\documentclass[sigplan,nonacm,10pt,balance=false]{acmart}
\usepackage{amsthm}
\usepackage{placeins}
\usepackage{stfloats}
\usepackage{pifont}

\newtheoremstyle{problemstyle}{6pt}{6pt}{\itshape}{}{\bfseries}{.}{.5em}{}
\theoremstyle{problemstyle}

\usepackage{booktabs}
\usepackage{colortbl}
\usepackage{enumitem}
\usepackage{multirow}
\usepackage{tabularx}
\newcolumntype{R}{>{\raggedleft\arraybackslash}X}
\usepackage{siunitx}
\usepackage{tikz}
\usetikzlibrary{arrows.meta,positioning,calc,decorations.pathreplacing,patterns}
\definecolor{tlblue}{RGB}{74,111,165}    %
\definecolor{tlgreen}{RGB}{91,165,133}   %
\definecolor{tlamber}{RGB}{217,149,59}   %
\definecolor{tlred}{RGB}{163,31,52}      %
\definecolor{tlnavy}{RGB}{12,35,64}      %
\definecolor{tlplum}{RGB}{122,92,144}    %
\colorlet{qubitfill}{tlplum!45}
\colorlet{magicfill}{tlgreen!50}
\colorlet{pathfill}{tlamber!75}
\colorlet{lanefill}{tlamber!22} %
\colorlet{freefill}{black!7}
\colorlet{tileedge}{white}
\colorlet{pathlabel}{white}
\newlength{\tilesize}
\DeclareRobustCommand{\circled}[2][tlred]{\tikz[baseline=(c.base)]\node[circle,fill=#1,text=white,inner sep=0.4pt,minimum size=1.2em,font=\scriptsize\bfseries](c){#2};}
\newcommand{\qsq}{\tikz[baseline=-0.3ex]\draw[fill=qubitfill,draw=tileedge,line width=0.7pt] (0,0) rectangle (0.7em,0.7em);}
\newcommand{\msq}{\tikz[baseline=-0.3ex]\draw[fill=magicfill,draw=tileedge,line width=0.7pt] (0,0) rectangle (0.7em,0.7em);}
\newcommand{\fsq}{\tikz[baseline=-0.3ex]\draw[fill=freefill,draw=tileedge,line width=0.7pt] (0,0) rectangle (0.7em,0.7em);}
\newcommand{\psq}{\tikz[baseline=-0.3ex]\draw[fill=pathfill,draw=tileedge,line width=0.7pt] (0,0) rectangle (0.7em,0.7em);}

\usepackage[ruled,vlined]{algorithm2e}
\usepackage{amsmath,amsfonts}
\usepackage[font=small,labelfont=bf]{caption}
\usepackage[capitalise,noabbrev]{cleveref}

\newcommand{\CX}{\ensuremath{\mathtt{CX}}}

\newcommand{\dascot}{DasCot\xspace}
\usepackage{xspace}

\newcommand{\headlineVolumeSaving}{55.6}

\newcommand{\headlineTopolsVolumeSaving}{61.9}
\newcommand{\headlineIdealApplications}{48}
\newcommand{\headlineIdealOverhead}{12.4}

\newcommand{\headlineDascotSpeedup}{75.5}
\newcommand{\headlineTopolsSpeedup}{2,240}

\newcommand{\headlineDascotUtilization}{9.5}
\newcommand{\headlineRVLaneUtilization}{28.0}
\newcommand{\headlineUtilizationImprovement}{18.5}
\newcommand{\sensitivityCollisionDefault}{6}
\newcommand{\sensitivityLaneDefault}{4}
\newcommand{\sensitivityPlacementDefault}{100}
\newcommand{\sensitivityBudgetDefault}{6}
\newcommand{\sensitivityVolumeDeviation}{49.6}

\begin{document}
\title[Demand-Driven Routing Space for Lattice-Surgery Compilation]{Demand-Driven Routing Space for Lattice-Surgery Compilation}

\author{Hanyu Wang}
\affiliation{%
  \institution{University of California, Los Angeles}
  \city{Los Angeles}
  \state{CA}
  \country{USA}}
\email{hanyuwang@g.ucla.edu}

\author{Wan-Hsuan Lin}
\affiliation{%
  \institution{QuEra Computing Inc.}
  \city{Boston}
  \state{MA}
  \country{USA}}
\email{WLin@quera.com}

\author{Zihan Chen}
\affiliation{%
  \institution{Rutgers University}
  \city{New Brunswick}
  \state{NJ}
  \country{USA}}
\email{zihan.chen.cs@rutgers.edu}

\author{Eddy Z. Zhang}
\affiliation{%
  \institution{Rutgers University}
  \city{New Brunswick}
  \state{NJ}
  \country{USA}}
\email{eddy.zhengzhang@gmail.com}

\author{Jason Cong}
\affiliation{%
  \institution{University of California, Los Angeles}
  \city{Los Angeles}
  \state{CA}
  \country{USA}}
\email{cong@cs.ucla.edu}

\begin{abstract}
Lattice-surgery compilation places logical qubits on a grid of surface-code tiles and routes their interactions.
Sparse static layouts limit routing contention but reserve space that often remains idle.
Reducing this reservation requires balancing area savings against additional gate serialization and qubit movement.
In this paper, we present a co-designed dynamic architecture and compiler that allocates routing space according to circuit demand.
The architecture preserves a fixed virtual qubit floorplan while allowing its physical spacing to change between execution stages.
The compiler coordinates placement, scheduling, and routing to implement lattice surgeries using fewer physical qubits.
Across our application benchmarks, our method reduces space--time volume by \headlineVolumeSaving\% in geometric mean and compiles $\headlineDascotSpeedup\times$ faster than the evaluated \dascot{} configuration.
A circuit-synthesis case study shows how our compiler exposes optimization tradeoffs.
\end{abstract}

\maketitle

\section{Introduction}
\label{sec:intro}

\suppressfloats[t]
\begin{figure}[!t]
  \centering

  \includegraphics[trim=0 2.8bp 0 1bp,clip]{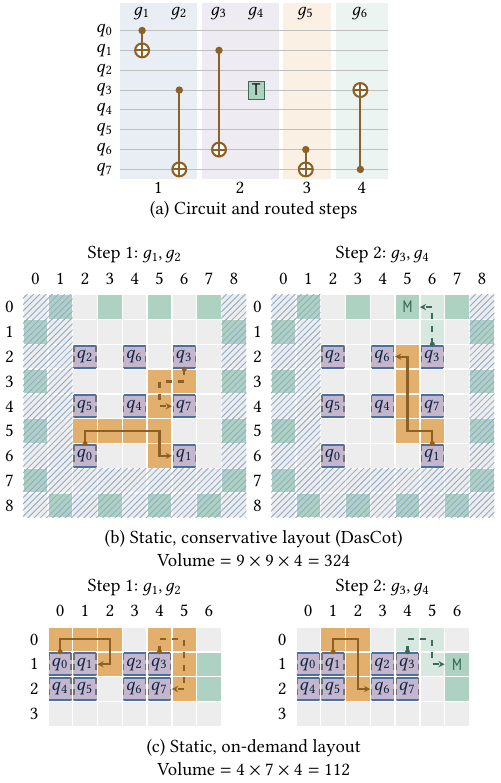}
  
  \caption{Motivating example: static, conservative layout (b) and static, on-demand layout (c).}
  \Description{Three vertically stacked panels show a six-gate circuit, two routing snapshots of the static nine-by-nine conservative layout, and two routing snapshots of the static four-by-seven on-demand layout with unused columns removed. Their volumes are 324 and 112 tile-steps, respectively. Four light background bands group the circuit gates by routed step. Each pair of snapshots shows steps one and two side by side. Two CX gates run together in step one; CX and T run together in step two. All path pairs in a snapshot are disjoint. Row and column indices in both layouts are zero-based.}
  \label{fig:example}
\end{figure}

\begin{figure}[!b]
  \centering

  \includegraphics[trim=0 2.8bp 0 0bp,clip]{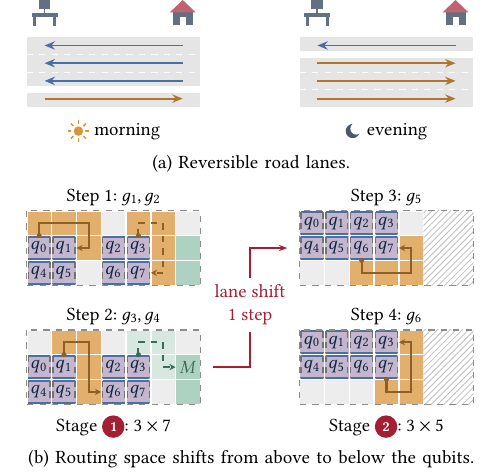}
  
  \caption{Reversible lanes (a) and our method with dynamic routing space (b).}
  \Description{Two stacked panels. Panel (a) places morning and evening road configurations side by side, with the sun and morning label beneath the left road and the moon and evening label beneath the right road. Three lanes lead toward work in the morning and toward home in the evening. Panel (b) aligns two routing stages below the roads. Stage one uses routing space above the qubits for steps one and two; stage two uses space below them for steps three and four. A one-step lane shift connects the stages on the same twenty-one-tile chip.}
  \label{fig:idea}
\end{figure}

Fault-tolerant quantum programs use many physical qubits to protect each logical qubit~\cite{gottesman1997stabilizer, fowler2012surface}.
Surface-code experiments have demonstrated improved logical memory as the code distance increases~\cite{google2023suppressing, google2025quantum}.
In the planar surface code, a logical qubit occupies a patch of physical qubits, represented here by a tile~\cite{horsman2012surface, litinski2019game}.
On a planar device, lattice surgery implements logical interactions by merging and splitting neighbouring patches~\cite{horsman2012surface, litinski2019game, fowler2018low, debeaudrap2020zx}.
Distant qubits therefore need intervening space for their interactions.
A compiler must determine both where logical qubits reside and how their interactions use the available space.
These decisions determine the required chip area and the number of execution steps, whose product is the space--time volume~\cite{gidney2021factor, gidney2025factor}.

Exposing this cost efficiently is important for algorithm--architecture co-design~\cite{mcardle2025fast}.
Analytical resource estimators predict execution cost without constructing the paths needed to establish routing feasibility~\cite{harrigan2024expressing, beverland2022assessing, huggins2025flasq}.
More detailed compilers construct three-dimensional realizations using SAT synthesis, topological transformations, or path search~\cite{tan2024sat, liao2026koval, topologiq2024, topols2025, hamada2024efficient}.
Nevertheless, minimizing lattice-surgery resource costs is computationally hard in general~\cite{herr2017optimization}. Therefore, compilers such as \dascot{}~\cite{dascot2024} and LibLSQECC~\cite{watkins2024high, leblond2023realistic, leblond2025comparison} use the patch-layer abstraction that represents execution as routing steps on a two-dimensional tile grid, enabling algorithm designers to efficiently compare circuit formulations.

However, routing feasibility can come at the cost of substantial idle space.
The \dascot{} layout evaluated here fixes data-qubit positions, separates neighbouring sites, and reserves a border around the array, as \cref{fig:example}b shows.
Each \CX{} or \texttt{T} gate uses a path of free tiles.
The reserved space limits contention, keeping routed step counts close to the depth imposed by circuit dependencies.
Yet only 5--27\% of the free tiles carry paths in an average step across our application benchmarks.
The resulting programs therefore reserve physical-qubit capacity that most execution steps leave idle.

Reducing chip area is an important yet challenging problem. 
Indeed, sharing lanes saves area but can introduce routing conflicts and serialize gates~\cite{litinski2019game}.
Conversely, routes that permit more gates to execute together may use different lanes and prevent the layout from being compacted.
Changing qubit placement can accommodate different routing demands, but movement adds execution time.
These tradeoffs make routing-space allocation an architecture--compiler co-design problem:
the architecture must expose opportunities to share and rearrange routing space, while the compiler must exploit those opportunities without excessive contention or movement.

\begingroup\tolerance=100\emergencystretch=1em
To address this problem, we present a co-designed dynamic architecture and lattice-surgery compiler.\footnote{An open-source implementation is available at \url{https://github.com/Nozidoali/MiniFlash}.}
Our architecture allows qubits to share routing lanes and their physical spacing to change between execution stages.
The compiler couples demand-driven floorplanning and contention-aware placement with scheduling, routing, and compaction.
Unlike previous works that route qubits after the mapping is given, we use traffic to dynamically rearrange qubits and allocate lanes.
Experimental results show that this dynamic execution model not only reduces the space-time volume by \headlineVolumeSaving\% but also speeds up the compilation by $\headlineDascotSpeedup\times$, compared with one of the state-of-the-art static models.
\par\endgroup

\section{Lanes Should Follow the Traffic!}
\label{sec:example}
Our method is inspired by reversible road lanes in \cref{fig:idea}a, with more lanes leading toward work in the morning and toward home in the evening.
In this section, we use the running example in \cref{fig:example}a to demonstrate the advantages of on-demand routing lane allocation and dynamic architecture, thereby motivating our method.
The circuit has eight qubits and six gates: five two-qubit \CX{} gates and one \texttt{T} gate.
The gates $g_1,g_3,g_5,g_6$ must execute in order, requiring four steps.
Gate $g_2$ can run alongside $g_1$, and $g_4$ alongside $g_3$.
In the snapshots, orange paths connect pairs of qubits, and the green path connects a qubit to an auxiliary state required by its \texttt{T} gate.
For the static layouts in \Cref{fig:example}, we specify the qubit placements for illustration and show only the first two routed steps.

\noindent\textbf{Static, conservative layout} (\cref{fig:example}b). A simple way to complete routing is to conservatively reserve space between neighbouring qubits and around the array~\cite{dascot2024}. This produces a $9\times9$ grid.
The routes within each of the two snapshots do not overlap, allowing both gates to run simultaneously.
All six gates finish in four steps, giving a space-time volume of $9\times9\times4=324$ tile-steps.
Nevertheless, as the hatching shows, many rows and columns remain unused throughout execution.

\noindent\textbf{Static, on-demand layout} (\cref{fig:example}c). One can reduce chip area by removing unused lanes, such as column 0 and row 8 in \cref{fig:example}b. A more systematic approach allocates routing space on demand, yielding the $4\times7$ layout shown in \cref{fig:example}c.
The volume falls to $4\times7\times4=112$ tile-steps without delaying any gate.
This saving does not require qubit movement.
However, as the execution depth grows, lanes are less likely to stay completely empty. All four rows must remain in this example; row 3, for instance, is needed in the last two steps.

\noindent\textbf{Dynamic routing space} (\cref{fig:idea}b). Based on this observation, our method adopts a dynamic architecture. By allowing different layouts at different stages, we can remove lanes that are unused within a partition of the entire circuit.
As shown in \cref{fig:idea}b, changing qubit spacing after the first two steps shifts routing space from above to below the qubits. The first stage uses a $3\times7$ layout, and the second uses $3\times5$.
This way, both layouts fit on one $3\times7$ chip.
This area reduction incurs a movement cost between stages. Recent developments in walking surface codes support parallel movement~\cite{mcewen2023relaxing}; in this example, under our movement model, the transition adds one execution step.
All four gate steps remain unchanged, so volume becomes $3\times7\times(2+1+2)=105$ tile-steps.

\section[Patch-Layer Execution Model]{Patch-Layer Execution Model}
\label{sec:formulation}

This section provides preliminary background on patch-layer execution.
Due to space limits, we refer readers to Horsman et al.~\cite{horsman2012surface} and Litinski~\cite{litinski2019game} for a more detailed and formal introduction.

\subsection{Circuit as a Traffic Dependency Graph}
\label{sec:form-execution}

We take a Clifford+\texttt{T} circuit with $n$ declared data qubits,
all resident throughout execution, and extract a directed acyclic
\emph{traffic dependency graph}~(TDG).
Each vertex represents an operation that requires routing:
a \CX{} between control and target qubits, or a
\texttt{T}/\texttt{T}$^\dagger$ gate between a data qubit
and a fresh magic state.
\Cref{fig:frontier-routing}a shows an example TDG.
Single-qubit Clifford gates such as \texttt{H}, \texttt{S},
\texttt{X}, and \texttt{Z} are not represented as routed
operations and are assigned no routing or execution cost
in our patch-layer model.
This modeling assumption omits Clifford implementation and
frame-management overhead, whose physical cost depends on the
chosen protocol~\cite{litinski2018twist, huggins2025flasq}.

\begin{figure}[tbp]
  \centering

  \includegraphics[trim=0 0.9bp 0 0bp,clip]{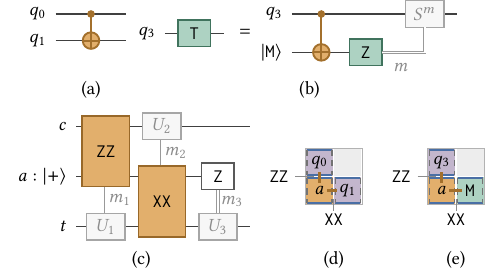}
  
  \caption{(a)~\CX{} and (b)~\texttt{T} with its injection circuit share (c)~the same \CX{} surgery~\cite{litinski2018twist}, with patch layouts in (d,e). Gray corrections are $U_1=\texttt{X}^{m_1}$, $U_2=\texttt{Z}^{m_2}$, and $U_3=\texttt{X}^{m_3}$, where $m_i=0$ denotes $+1$ parity.}
  \Description{A single-column figure with two rows. The top row shows a vertically compact CX in panel a and a single T gate on q3 followed by an equals sign and its injection circuit in panel b, with the injection's magic-state readout and conditional S still visible. The bottom row shows their common CX surgery followed by two compact patch layouts. The surgery's control c and target t interact through a plus-state ancilla a, using successive merge/split measurements of ZZ then XX and ancilla Z readout. Each measurement is aligned with a gray correction on another wire and joined to it by a double classical line. U1 and U3 correct the target, and U2 the control. The two-by-two tile grids on the right assign c,t to q0,q1 or q3,M. Both routes pass through the bottom-left ancilla, with the top-right tile unused.}
  \label{fig:gate-surgery}
\end{figure}

\textbf{Vertices:}
Both types of routed operations require connectivity between patches,
as illustrated in \cref{fig:gate-surgery}.
For a \CX{} (\cref{fig:gate-surgery}a), a temporary ancilla patch is initialized to $|+\rangle$.
Two successive merge/split operations measure its joint \texttt{ZZ} parity
with the control and \texttt{XX} parity with the target, before the ancilla
is read out~\cite{litinski2018twist}.
A \texttt{T} gate (\cref{fig:gate-surgery}b) consumes a fresh magic state
$|\texttt{M}\rangle=(|0\rangle+e^{i\pi/4}|1\rangle)/\sqrt{2}$
through a \CX{} and readout, with an outcome-dependent
\texttt{S} correction~\cite{bravyi2005universal}.
It therefore requires traffic from the data qubit
to any magic-state patch.
We count each gate's ordered surgery operations as one
\emph{routed step}, or simply \emph{step}.

\textbf{Edges:}
An edge $g_i\to g_j$ requires $g_i$ to finish before $g_j$ starts.
The dependency depth $D$ is the number of gates on a longest TDG
chain and is a lower bound on the number of routed steps.
The TDG may leave commuting gates unordered.
For example, \CX{} gates sharing only a control or only a target
can execute in either order, provided their dependencies on other
gates are respected.
Commutation permits reordering, but does not by itself permit
concurrent execution: gates sharing a data tile must still execute
in different routed steps.
Likewise, separate magic-state consumptions introduce no circuit
dependency between otherwise independent gates, but remain subject
to magic-state availability.

\subsection[Patch-Layer Abstraction]{Patch-Layer Abstraction}
\label{sec:form-output}

A \emph{tile connectivity graph}~(TCG) records qubit placements
and the tile connectivity allocated to gates in one routed step,
such as the first step of \cref{fig:example}b.
The output program consists of a sequence of TCGs.
All gate steps use the same fixed grid of $H$ rows
and $W$ columns.

\begin{figure}[!t]
  \centering

  \includegraphics[trim=0 0.4bp 0 0.5bp,clip]{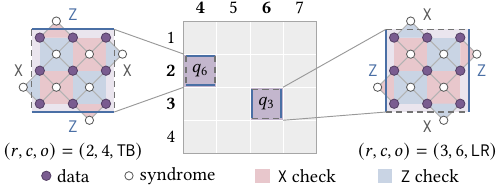}
  
  \caption{Distance-3 patches represented by tiles in two orientations~\cite{litinski2019game}.}
  \Description{A central four-by-four grid highlights q6 at row two, column four, with orientation TB, and q3 at row three, column six, with orientation LR. Two zoom lines connect each tile to an enlarged distance-three patch on its side of the grid. Each patch contains nine physical data qubits and eight syndrome qubits. The left patch has solid blue Z boundaries above and below and dashed gray X boundaries on its sides. The right patch, including its syndrome qubits, is rotated ninety degrees, so these boundary directions are exchanged. Red and blue filled regions denote X and Z checks, respectively, with square interior checks and triangular boundary checks. The qubit placements and orientations are illustrative.}
  \label{fig:patch-tile}
\end{figure}

\noindent\textbf{Qubit placement.}
A \emph{qubit placement} is a tuple $(r,c,o)$, where $r$ and $c$
are zero-based tile coordinates and
$o\in\{\texttt{TB},\texttt{LR}\}$ is the \emph{orientation}.
The labels name the sides carrying \texttt{Z} boundaries: top and bottom
for \texttt{TB}, or left and right for \texttt{LR}.
The remaining two sides carry \texttt{X} boundaries.
\Cref{fig:patch-tile} illustrates both orientations using
distance-3 surface-code patches.

\noindent\textbf{Traffic routing.}
In each TCG, a vertex denotes a data-qubit tile, a routing tile
used in that step, or a supplied magic-state tile.
Unused routing tiles are omitted from the graph, but remain
part of the charged chip area.
An undirected edge connects adjacent tiles participating in a
gate's lattice-surgery operations, abstracting two consecutive merge/splits. Figures in this paper include arrows just to differentiate control and target of CX, because they must connect to different types of boundaries. Besides,
routes executed in the same step must be tile-disjoint,
including their data-qubit and magic-state endpoints.

For example, in the first step of \cref{fig:example}b,
$g_2$ connects $q_3$ at $(2,6)$ to $q_7$ at $(4,6)$ through
routing tiles $(3,6)$, $(3,5)$, and $(4,5)$.
These five vertices and the edges joining consecutive tiles
form one connected component.

\section[Dynamic Architecture]{Dynamic Architecture}
\label{sec:architecture}

Dynamic routing-space allocation couples architecture design
with compilation.
Using less routing space reduces area, but can introduce contention
and serialize gates.
Changing qubit spacing can accommodate different routing demands,
but movement adds execution time.
This section specifies the dynamic architecture constraints and formulates the space--time volume objective.

\subsection[Virtual Floorplan and Dynamic Mapping]{Virtual Floorplan and Dynamic Mapping}
\label{sec:form-lanes}
\label{sec:form-rearrange}

As shown in \cref{sec:example}, dynamic qubit placement allows routing space to adapt to changing circuit traffic.
Ideally, qubits could be rearranged arbitrarily between steps, but this could introduce significant movement overhead.
To strike a better balance, our method draws on walking surface codes for parallel movement~\cite{mcewen2023relaxing} and separates qubit assignment from physical spacing.
A \emph{virtual floorplan} fixes the array shape, qubit-to-site
assignment, and qubit orientations throughout execution.
Its \emph{physical mapping} determines the spacing between
qubit rows and columns and may change between stages.

\begin{figure}[tbp]
  \centering

  \includegraphics[trim=0 2.8bp 0 2.8bp,clip]{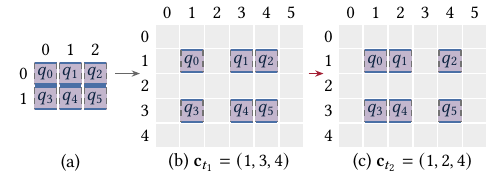}
  
  \caption{Two different lane spacings in (b) and (c) for the same virtual floorplan in (a). The transition from (b) to (c) moves the middle qubit column one tile left, giving $\delta_s=1$.}
  \Description{The same six qubits keep their virtual row and column order. Two physical layouts place the virtual columns at coordinates 1, 3, 4 and 1, 2, 4, respectively. Both place the virtual rows at 1 and 3. Only the middle qubit column moves.}
  \label{fig:case-lanes}
\end{figure}

More specifically, $\pi(q_i)=(u_i,v_i,o_i)$ assigns qubit $q_i$
a distinct site $(u_i,v_i)$ in an $h\times w$ virtual array
and a fixed orientation $o_i$, with $hw\ge n$.
At execution step $t$, a row-coordinate vector $\mathbf{r}_t$
of length $h$ and a column-coordinate vector $\mathbf{c}_t$
of length $w$ map virtual indices to physical tile coordinates:
\begin{equation}
  p_t(q_i)=\bigl(\mathbf{r}_t[u_i],\mathbf{c}_t[v_i],o_i\bigr).
  \label{eq:dynamic-mapping}
\end{equation}
Both vectors contain strictly increasing integer coordinates
within the fixed chip.
Changing an entry moves the qubits in the corresponding virtual
row or column together, while preserving their order.
Gaps between mapped rows and columns determine routing-lane widths.
In \cref{fig:case-lanes}, changing $\mathbf{c}_t$ from $(1,3,4)$
to $(1,2,4)$ redistributes the internal vertical routing space
without changing the virtual floorplan.

Execution is partitioned into $K$ \emph{stages}, each a consecutive group of routed steps that uses one physical mapping.
Walks between stages must preserve row and column order and remain within the fixed chip.

\subsection{Objective Function}
\label{sec:form-objective}

We minimize the \emph{space--time volume}: the fixed grid area
multiplied by the total number of execution steps.

Let $H_s$ and $W_s$ denote stage $s$'s dimensions, including
space for magic-state delivery, and let $T_s$ be its number
of routed steps.
The fixed chip must accommodate every stage, while total
execution time includes both gate steps and walks.
The volume is therefore
\begin{equation}
  \begin{aligned}
    H&=\max_{1\le s\le K}H_s,\qquad W=\max_{1\le s\le K}W_s,\\
    \mathrm{Volume}&=HW\left(\sum_{s=1}^{K}T_s+\sum_{s=1}^{K-1}\delta_s\right).
  \end{aligned}
  \label{eq:grid}
\end{equation}
\begin{figure}[t]
  \centering

  \includegraphics{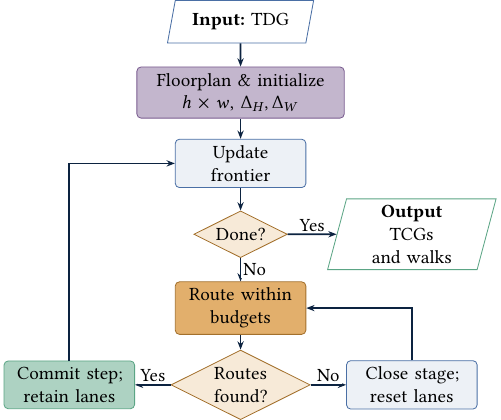}
  
  \caption{Main compiler workflow of our method, from the input TDG to compacted output TCGs and walks. Selected routes determine each stage's mapping. Closing a stage resets lane usage and retries the same frontier.}
  \Description{The input TDG feeds the compiler loop, which starts by choosing the virtual floorplan dimensions h by w and initializing the row and column lane budgets Delta H and Delta W. While traffic remains, routing selects a compatible step under the stage budgets. A selected step is committed and the frontier is updated. Otherwise, the current stage closes and routing resumes with fresh lane usage. The output box contains the compacted TCGs and walks.}
  \label{fig:compiler-workflow}
\end{figure}

Different stages may determine the maximum height and width.
Padding around smaller stages is charged at every step.

Let $(r_i^{(s)},c_i^{(s)})$ denote the physical coordinates of
qubit $q_i$ in stage $s$.
We model the duration $\delta_s$ of the walk after stage $s$ as the largest
absolute row or column displacement of any qubit:
\begin{equation}
  \delta_s=\max_{0\le i<n}\max\left\{
    \left|r_i^{(s+1)}-r_i^{(s)}\right|,
    \left|c_i^{(s+1)}-c_i^{(s)}\right|
  \right\},
  \label{eq:walk-duration}
\end{equation}
for $1\le s<K$.
The walk contributes $\delta_s$ steps to execution time.
For the transition from \cref{fig:case-lanes}b to
\cref{fig:case-lanes}c, only $q_1$ and $q_4$ move one tile
left, giving $\delta_s=1$.
The one-step-per-tile cost is based on the recently developed walking surface code scheme, consistent with the optimistic movement estimate in
FLASQ~\cite{huggins2025flasq}.

\section[Demand-Driven Compiler]{Demand-Driven Compiler}
\label{sec:method}

\begingroup\emergencystretch=0.5em
The dynamic architecture requires scheduling, routing, and mapping to account for how their choices affect space--time volume.
Scheduling must balance dependency progress against the routing space retained by each stage; routing should favor lane reuse; and mapping must anticipate both contention and compaction.
The compiler coordinates these choices with stage formation to balance area savings against gate delays and movement overhead.
\par\endgroup

Exact joint optimization of placement, routing, and stage boundaries could fully exploit the architecture's dynamism, but greatly enlarges the compilation search space.
Searching this larger space can even yield worse solutions when compilation time is bounded.
To balance solution quality and compile time, our method fixes a virtual floorplan, routes under lane budgets, and compacts each stage according to its routing demand.
\Cref{fig:compiler-workflow} shows this workflow.
We first introduce shape selection and lane budgets in \cref{sec:layout}, then frontier routing in \cref{sec:routing}, followed by qubit placement in \cref{sec:mapping}.

\subsection[Demand-Driven Floorplanning]{Demand-Driven Floorplanning}
\label{sec:layout}
\label{sec:dynamic-mapping}
\label{sec:shift}
\label{sec:form-stages}
\label{sec:form-demand}
\label{sec:lane-routing}

Floorplanning selects the virtual array's shape and the routing space available within each stage.
We first predict a shape from circuit demand, then bound stage dimensions with row and column lane budgets.

\noindent\textbf{Shape selection.}
Since the input fixes the number of qubits $n$, we choose the virtual array's aspect ratio.
We consider $h=1,\ldots,\lceil\sqrt n\rceil$ site rows and
$w=\lceil n/h\rceil$ columns, selecting the shape with minimum predicted volume:
\[
  (h^\star,w^\star)
    =\underset{\substack{1\le h\le\lceil\sqrt n\rceil\\
                         w=\lceil n/h\rceil}}{\arg\min}
      \widehat{V}(h,w).
\]

\begin{figure}[!b]
  \centering

  \includegraphics{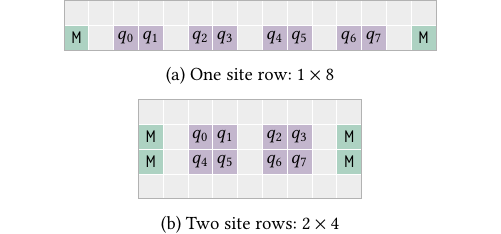}
  
  \caption{Eight qubits in two shapes. The shape model balances initial grid area against magic-state delivery and routing capacity.}
  \Description{The one-by-eight floorplan has eight purple qubit tiles, two green magic-delivery tiles, and twenty gray routing tiles, in a two-by-fifteen grid. The two-by-four floorplan has the same eight qubits, four magic-delivery tiles, and twenty-four routing tiles, in a four-by-nine grid.}
  \label{fig:shape-demand}
\end{figure}

To estimate $\widehat{V}$, we use a throughput model inspired by Roofline~\cite{williams2009roofline}.
Depending on the TDG's gate demand and the chosen aspect ratio, execution can be limited by magic-state delivery or \CX{} routing.
\begin{itemize}[nosep]
  \item \textbf{Magic and \CX{} demand.}
  We group gates at their earliest steps allowed by the TDG.
  Let $m_{\mathrm{CX},j}$ and $m_{\mathrm{T},j}$ count the \CX{} and
  \texttt{T}/\texttt{T}$^\dagger$ gates in group $j$, respectively.

  \item \textbf{Magic and \CX{} resources.}
  Although lane allocation is determined during routing, we estimate throughput using a proxy:
  two \texttt{T} gates per site row per step, supplied by the two magic-delivery tiles at its ends,
  and one \CX{} route per site row per step.
  As illustrated in \cref{fig:shape-demand}, this gives estimated throughputs of $2h$ \texttt{T} gates and $h$ \CX{} gates per step.
\end{itemize}

Putting these together, a simplified volume estimate using the virtual area $h\times w$ is

\begin{equation}
    \widehat{V}(h,w)
      =h\times w\times\sum_{j=1}^D\max\left\{
        1,
        \left\lceil\frac{m_{\mathrm{CX},j}}{h}\right\rceil,
        \left\lceil\frac{m_{\mathrm{T},j}}{2h}\right\rceil
      \right\}.
  \label{eq:predicted-volume}
\end{equation}
The maximum selects the bottleneck, with at least one step per group.
If neither throughput limits execution, the estimate reduces to $D\times h\times w$.
The single-row shape attains the minimum virtual area $h\times w=n$; ties favor fewer site rows.

Our shape-selection heuristic does not guarantee feasibility:
it omits actual routing conflicts, compaction, and walks, and \texttt{T} traffic also competes for routing space.
We evaluate its effectiveness against exhaustive configuration selection in \cref{sec:eval-full}.

\noindent\textbf{Lane budgets within a stage.}
For an $h\times w$ virtual floorplan, we next determine lane budgets
$\Delta_H,\Delta_W$ that limit the additional rows and columns retained
for routing and magic-state delivery:
\begin{equation}
  H_s\le B_H=h+\Delta_H,\qquad
  W_s\le B_W=w+\Delta_W.
  \label{eq:lane-budgets}
\end{equation}

\begin{figure}[tbp]
  \centering

  \includegraphics[trim=0 0bp 0 0.4bp,clip]{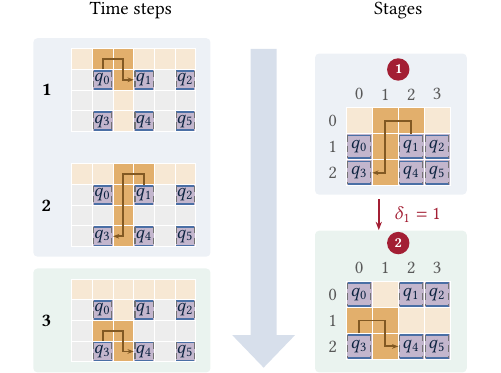}
  
  \caption{On-demand spacing with lane budgets $\Delta_H=\Delta_W=1$. Left: steps 1--2 share one lane row and column; step 3 needs another row and starts stage \circled{2}. Right: compacted $3\times4$ stages, showing their last routing steps, with a one-step walk. Backgrounds distinguish stages; pale orange marks retained lanes, darker orange routes.}
  \Description{Three routing snapshots are stacked vertically on the left, labeled one, two, and three at their left edges. The left grids omit row and column coordinates. Steps one and two reuse one lane row and one lane column and are grouped into stage one. Step three requests a second row, so stage one closes and lane usage resets. Matching pale-blue backgrounds group steps one and two with stage one; pale-green backgrounds group step three with stage two. On the right, compacted stage one is aligned with the first two steps, and compacted stage two with step three. Both have three rows and four columns and show physical tile coordinates. A broad light-blue downward arrow separates the columns and indicates time progression. A small crimson arrow between stages denotes a one-step walk. Routing step numbers exclude that walk. Pale orange marks allocated lanes and darker orange marks the shown routes.}
  \label{fig:route-cands}
\end{figure}

Each stage initially retains only data-site coordinates.
As illustrated in \cref{fig:route-cands}, we route successive steps under these budgets using the selection rule in \cref{sec:routing},
accumulating required coordinates and updating the ready gates after each step.
If no route is selected in a nonempty stage, we close it and retry
the same ready gates with fresh lane resources.

Lane budgets and routing choices are interdependent:
routes determine lane demand, while budgets guide route selection and stage boundaries.
We fix the budgets before routing to simplify route selection and stage partitioning (\cref{sec:routing}).

\subsection[Compaction-Aware Frontier Routing]{Compaction-Aware Frontier Routing}
\label{sec:form-scheduling}
\label{sec:routing}

At routed step $t$, the \emph{dependency frontier} $\mathcal{F}_t$
contains unexecuted TDG gates whose predecessors have finished.
The router selects a subset of frontier gates and one candidate
path for each selected gate.
Routes connect control \texttt{Z} to target \texttt{X} boundaries for \CX{},
or a data \texttt{Z} boundary to a fresh magic-state tile for \texttt{T}.
Selected routes cannot share data, routing, or consumed magic-state
tiles.
After committing a routed step, the compiler updates the frontier.

Routing algorithm should not simply maximize concurrency in the current step, as it does not necessarily
minimize program volume. Indeed,
executing several short dependency branches may delay a gate on a
longer remaining chain.
Likewise, routes spread across different lanes can prevent a stage
from shrinking, even when they allow more gates to execute together.

Instead, we select routes using dependency criticality, the additional
coordinates retained after compaction, and row and column budgets.
For each frontier gate, we search paths in increasing order of bend count and retain at most one candidate.
For \texttt{T}/\texttt{T}$^\dagger$ gates, the search is restricted to the nearest available magic-state tiles.
Compared with a general shortest-path search, this restriction reduces search effort and favors low-bend routes, but may miss feasible routes.

\noindent\textbf{Compatibility and criticality.}
The candidates define an undirected \emph{conflict graph}
$\mathcal{C}_t$.
Each vertex represents a candidate route.
An edge joins two vertices when their routes share a data, routing, or consumed magic-state tile.
An independent set therefore selects mutually compatible routes.

We assign each gate $g$ a \emph{criticality} $\kappa_g$, the length
of the longest remaining TDG chain starting at $g$, including $g$
itself.
Writing $\mathrm{succ}(g)$ for its immediate successors,
\begin{equation}
  \kappa_g=
  \begin{cases}
    1, & \mathrm{succ}(g)=\emptyset,\\
    1+\max_{g'\in\mathrm{succ}(g)}\kappa_{g'}, & \text{otherwise}.
  \end{cases}
  \label{eq:criticality}
\end{equation}
Without lane costs or budgets, maximizing the total criticality
of compatible routes is a maximum-weight independent-set problem.

\noindent\textbf{Retained-coordinate cost.}
To account for lane demand, we penalize the additional coordinates
required by the selected routes.
For an independent set $I$ in the open stage $s$, we maximize
\begin{equation}
  \mathrm{score}_s(I)=\sum_{p\in I}\kappa_{g(p)}
    -\lambda\ell_s(I),
  \label{eq:route-score}
\end{equation}
subject to separate budgets on retained rows and columns.
Here, $g(p)$ is the gate implemented by route $p$, and $\ell_s(I)$ counts the additional row and column coordinates
that $I$ requires compaction to retain, beyond the data-site
coordinates and those already retained in stage $s$.
Rows and columns count separately, and a coordinate shared by
several routes is counted only once.
Required coordinates come from route-segment endpoints and turns, and consumed magic-state tiles.
The nonnegative weight $\lambda$ encourages selected routes to reuse
coordinates within a step and across successive steps of the stage.
The objective thus extends criticality-weighted independent-set
selection with costs for retained coordinates and lane budgets.
Our router uses branch and bound to search for a compatible set of candidate routes that maximizes the score in \cref{eq:route-score} while respecting the row and column budgets.

\begin{figure}[tbp]
  \centering

  \includegraphics[trim=0 2.8bp 0 0bp,clip]{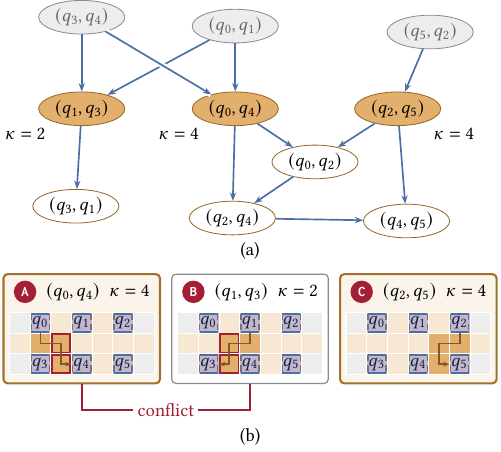}
  
  \caption{Frontier routing. (a) TDG: gray executed, orange ready, white blocked; $\kappa$ denotes criticality. (b) Conflict graph: pale orange marks retained lanes, darker orange routes.}
  \Description{Three completed gates precede a frontier of (q0,q4), (q1,q3), and (q2,q5), with criticalities 4, 2, and 4. Four further gates remain blocked. Each candidate route occupies three routing tiles on a three-row seven-column grid containing six data qubits. The first two routes share two tiles, and the third route is disjoint from both. Each route grid is enclosed in a box that acts as a conflict-graph vertex. Pale orange free tiles mark retained row one and columns two and four; the candidate routes are darker orange. The boxes are labeled A, B, and C from left to right. A red edge labeled conflict joins A and B. Amber frames select A and C.}
  \label{fig:frontier-routing}
\end{figure}

In \cref{fig:frontier-routing}, candidates \circled{\texttt{A}}, \circled{\texttt{B}},
and \circled{\texttt{C}} have criticalities 4, 2, and 4.
\texttt{A} conflicts with \texttt{B}, while \texttt{C} is compatible
with both.
All three reuse retained coordinates, so the lane penalty is zero.
We select \texttt{A} and \texttt{C}, with total criticality 8,
rather than \texttt{B} and \texttt{C}, with total criticality 6,
and defer \texttt{B} to a later step.

\begin{figure}[tb]
  \centering

  \includegraphics[trim=0 2.8bp 0 0.5bp,clip]{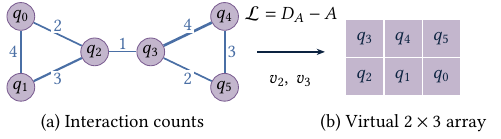}
  
  \caption{Spectral initialization on six qubits.}
  \Description{A weighted graph arranges two strongly connected triples in an infinity-like pattern, with q2 and q3 at the center joined by a short bridge and horizontal edge labels. The native spectral initialization assigns one triple to each row of a two by three virtual array.}
  \label{fig:spectral-placement}
\end{figure}

\subsection[Qubit Placement]{Qubit Placement}
\label{sec:mapping}

Placement assigns qubits and orientations to sites in the chosen virtual floorplan, anticipating contention and routing-space demand.

\noindent\textbf{Initial placement.}
After selecting the shape, we place frequently interacting qubits
near one another using \emph{spectral initialization}.
We construct a weighted qubit-interaction graph in which each edge
counts the \CX{} operations between its qubits.
A uniform weight of $10^{-3}$ between distinct qubits connects
otherwise disconnected components.
The second eigenvector of the graph Laplacian, the \emph{Fiedler
vector}, orders qubits into groups of at most $w$ sites, one per
virtual row.
Within each row, the third eigenvector orders qubits into
columns~\cite{hall1970rdimensional, fiedler1973algebraic}.
\Cref{fig:spectral-placement} illustrates this assignment.

\noindent\textbf{Placement refinement.}
We refine the site assignment and qubit orientations using simulated
annealing~\cite{Sechen1985TimberWolf}.
Let $\widehat{\chi}(\pi)$, $\widehat{\ell}(\pi)$, and
$\widehat{L}(\pi)$ estimate route overlaps, retained routing lines,
and path length, respectively.
We minimize the estimated placement cost
\begin{equation}
  \widehat{J}(\pi)=
  w_\chi\widehat{\chi}(\pi)
  +w_\ell\widehat{\ell}(\pi)
  +w_P\widehat{L}(\pi),
  \label{eq:placement-cost}
\end{equation}
The weights balance routing contention, retained routing space, and path length.
Overlaps can delay gates, while scattered paths keep more rows
and columns from being removed.

For each gate, a \emph{geometric proxy} connects its endpoint sides
through nearby lanes, using row-end magic for \texttt{T}.
These paths ignore occupancy and need not be legal routes.
For each dependency group, the retained-line estimate counts distinct row and column coordinates used by proxy endpoints, turns, and magic-state tiles, excluding data-site coordinates, then sums these counts across groups.
For the overlap estimate, each tile records its most recent route
within a dependency group, and a new route counts each encountered
previous owner once.
This scan can miss some pairwise overlaps when several routes
share a tile.

\begin{figure}[!t]
  \centering

  \includegraphics[trim=0 2.8bp 0 2.8bp,clip]{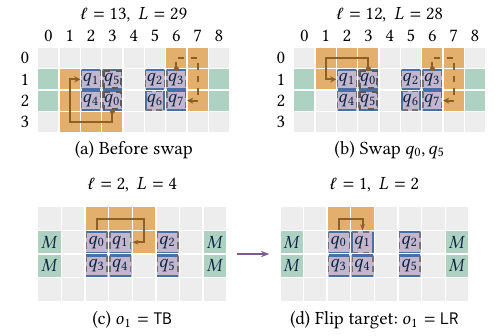}
  
  \caption{Proposals: (a--b) swap $q_0,q_5$; (c--d) flip $q_1$'s orientation.}
  \Description{Four panels in two rows. The top row shows site reassignment: swapping qubits zero and five reduces estimated retained lines from thirteen to twelve and path length from twenty-nine to twenty-eight. The bottom row shows an orientation flip of target qubit one: the same interaction uses a straight two-tile path instead of a bent four-tile path, reducing additional retained coordinates from two to one.}
  \label{fig:placement-moves}
  \label{fig:map-cases}
  \label{fig:orientation-move}
\end{figure}

The search uses two types of compile-time proposals:
\begin{itemize}
  \item \textbf{Site reassignment} (\cref{fig:placement-moves}a--b).
  Select a qubit and a destination site, swapping occupants if
  the site is occupied or moving the qubit if it is empty.
  In the example, $q_0$ and $q_5$ exchange sites, allowing $g_1$
  and $g_2$ to share the upper lane without overlapping.
  With orientations fixed, the whole-circuit line estimate falls
  from 13 to 12 and path length from 29 to 28, with no overlaps.

  \item \textbf{Orientation flip} (\cref{fig:placement-moves}c--d).
  Keep the site fixed and exchange the qubit's \texttt{Z} and \texttt{X}
  boundary directions.
  Flipping target $q_1$ lets $(q_0,q_1)$ use a straight route
  in the upper lane, shortening it from four tiles to two
  and requiring one additional retained coordinate rather than two.
  A flip may make other interactions more expensive, so its score
  includes every affected dependency group.
\end{itemize}

By default, 20\% of proposals flip orientations and 80\% reassign
sites.
Annealing accepts improvements and occasionally accepts higher costs
to escape local minima.
Larger cost increases and geometric cooling make these uphill moves
less likely.
Our method compiles the chosen floorplan once, and the final qubit assignment and orientations remain fixed throughout execution.

\section{Evaluation}
\label{sec:eval}

We evaluate how demand-driven routing-space allocation affects
space--time volume, compilation time, and routing-space utilization.
After describing the experimental setup (\cref{sec:eval-setup}), we compare volume and compilation time with the baselines
(\cref{sec:eval-full,sec:eval-time}), then analyze the area--time
tradeoff behind the volume savings (\cref{sec:eval-breakdown}).
A cumulative ablation examines the contributions of the compiler
components (\cref{sec:eval-ablation}), and a sensitivity study
examines the effects of cost weights and placement effort
(\cref{sec:eval-sensitivity-main}).

\subsection[Experimental Setup]{Experimental Setup}
\label{sec:eval-setup}

We perform experiments on 60 circuits from different application families, as summarized in \cref{tab:workloads}.
The suite includes arithmetic circuits, quantum Fourier transforms and phase estimation, QAOA, oracles, and eight random stress circuits.
These circuits vary in qubit count, \CX{} and \texttt{T} gate count, dependency depth, \texttt{T}-gate share, and interaction structure, from nearest-neighbour chains to all-to-all communication.
We run every compiler, our method, its solver controls, and the baselines, on the same AMD EPYC 9654 compute nodes, one CPU thread per compilation.
\begin{table*}[!t]
  \caption{The 60 benchmark circuits by family, with circuit counts and ranges of qubit count $n$, \CX{} and \texttt{T} gate count $|\mathcal{C}|$, and dependency depth $D$ (longest TDG chain in gates). The \texttt{T} share is the fraction of counted gates that are \texttt{T} gates.}
  \label{tab:workloads}
  \centering\small
  \setlength{\tabcolsep}{7pt}
  \begin{tabularx}{\textwidth}{@{}>{\raggedright\arraybackslash}Xrrrrr>{\raggedright\arraybackslash}X@{}}
    \toprule
    \rowcolor{tlplum!15}
    family & \# & $n$ & $|\mathcal{C}|$ & $D$ & \texttt{T} share & interaction \\
    \midrule
    \rowcolor{tlplum!4}
    Arithmetic & 28 & 5--192 & 50--53,437 & 24--2,314 & 33--64\% & all-to-all \\
    QFT, QPE & 5 & 4--18 & 93--1,335 & 58--443 & 65--87\% & T-heavy \\
    \rowcolor{tlplum!4}
    QAOA & 4 & 12--24 & 132--332 & 44--89 & 40--45\% & graph \\
    Random dense & 4 & 8--32 & 150--600 & 43 & 33\% & all-to-all \\
    \rowcolor{tlplum!4}
    Random sparse & 4 & 8--32 & 160--640 & 40 & 0\% & disjoint groups \\
    Oracles~\cite{bernstein1997quantum, deutsch1992rapid} & 7 & 8--30 & 7--39 & 7--29 & 0\% & sparse, serial \\
    \rowcolor{tlplum!4}
    Chains, FTCB kernels~\cite{harkness2026ftcircuitbench} & 8 & 4--200 & 121--34,691 & 87--19,345 & 0--91\% & nearest-neighbour \\
    \bottomrule
  \end{tabularx}
\end{table*}

\begin{figure*}[!b]
  \centering
  \includegraphics[width=\textwidth,trim=0 2.3bp 0 5.6bp,clip]{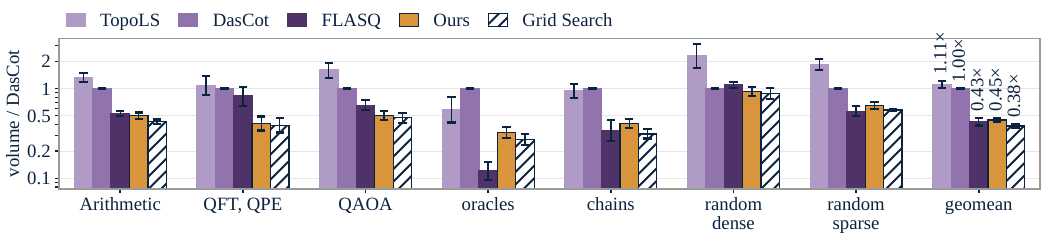}
  \caption{Geometric-mean volume relative to \dascot, with standard errors.}
  \Description{Grouped bars compare our method and a verified exhaustive Grid Search with DasCot, TopoLS, and FLASQ on a logarithmic vertical axis.}
  \label{fig:volume}
\end{figure*}

We compare our method with the following baselines:
\begin{itemize}[nosep,leftmargin=*]
  \item \textbf{\dascot}~\cite{dascot2024} uses the same tile-step volume definition as our method, although its endpoint and magic-delivery rules differ.
  \item \textbf{TopoLS}~\cite{topols2025} measures volume in tile-by-$d$-round cubes, where $d$ is the code distance; one \CX{} step of our method spans two such cubes in time.
  \item \textbf{FLASQ}~\cite{huggins2025flasq} provides an analytical resource estimate in blocks without constructing routes. We use Qualtran's implementation with row-major square placement and choose the device size to minimize volume.
\end{itemize}

Moreover, we compare with \emph{Grid Search}, which exhaustively enumerates the following hyperparameters for each $n$-qubit circuit:
\begin{itemize}[nosep,leftmargin=*]
  \item \textbf{Virtual layout shape:} $(h,w)$.
  \item \textbf{Extra-row budget:} $\Delta_H$.
  \item \textbf{Extra-column budget:} $\Delta_W$.
\end{itemize}
Grid Search selects the smallest verified volume after routing, compaction, and walks, holding the placement and routing algorithms and other settings fixed.
Our method instead uses the demand-based heuristic in \cref{sec:layout} to predict one shape.
Grid Search measures the benefit of exhaustive parameter selection within this family, rather than a global optimum.

For the main volume and ablation comparisons, we use one completed run per circuit for our method and its solver controls, with random seed 0.
We report geometric means across circuits, with standard errors computed in log space.

\subsection{Space--Time Volume Comparison}
\label{sec:eval-full}
\label{sec:eval-models}

\Cref{fig:volume} compares space--time volume across circuit families,
normalized to \dascot{}.
Although the improvement varies across families, our method achieves
lower geometric-mean volume in every family and improves most
individual applications.
In geometric mean, our method reduces volume by \headlineVolumeSaving\% relative to \dascot{} and reports \headlineTopolsVolumeSaving\% lower numerical volume than TopoLS under their respective cost models (\cref{sec:eval-setup}).
These results support allocating routing space according to circuit demand.

Our method also reports numerical volumes close to FLASQ's analytical
estimates. However,
FLASQ only estimates resources without constructing routes, whereas
Our method produces an explicit tile-level program with routes and
modeled movements under the architecture model in
\cref{sec:architecture}.
This distinction is important when interpreting the comparison:
numerically similar volumes under different cost models do not
establish equivalent physical resource requirements.

Grid Search measures the volume gap between our heuristic's configuration and the best shape and lane budgets within the searched family, using the same placement and routing algorithms.
Across the \headlineIdealApplications{} applications with completed searches, our method's volume is only \headlineIdealOverhead\% higher in geometric mean.
Grid Search obtains this improvement through repeated compilation across shapes and lane budgets.

\newcommand{\archCircuitCount}{10}
\newcommand{\archAreaImprovedCount}{10}
\newcommand{\archAreaUnchangedCount}{0}
\newcommand{\archAreaCompact}{0.96}
\newcommand{\archStepsCompact}{1.00}
\newcommand{\archVolumeCompact}{0.96}
\newcommand{\archAreaPaired}{0.60}
\newcommand{\archStepsPaired}{1.03}
\newcommand{\archVolumePaired}{0.61}
\newcommand{\archAreaRVLane}{0.44}
\newcommand{\archStepsRVLane}{1.13}
\newcommand{\archVolumeRVLane}{0.49}
\begin{table*}[!t]
  \caption{Architecture and compiler comparison. DasCot Compacted removes rows and columns unused throughout the original DasCot execution, preserving its step count. Area is in tiles and volume in tile-steps; steps include movement. Bold marks row minima, including ties. Geometric-mean ratios are relative to Static (DasCot).}
  \label{tab:architecture}
  \centering\footnotesize
  \setlength{\tabcolsep}{2.4pt}
  \begin{tabularx}{\textwidth}{@{}>{\raggedright\arraybackslash}Xrrrrrrrrrrrrrrrrrr@{}}
    \toprule
    & & & \multicolumn{4}{>{\columncolor{tlplum!15}}c}{\circled[tlplum]{\texttt{A}} Static (DasCot)} & \multicolumn{4}{>{\columncolor{tlplum!15}}c}{\circled[tlplum]{\texttt{B}} Static (DasCot Compacted)} & \multicolumn{4}{>{\columncolor{tlplum!15}}c}{\circled[tlplum]{\texttt{C}} Static (on-demand)} & \multicolumn{4}{>{\columncolor{tlamber!18}}c}{\circled[tlamber]{\texttt{D}} Ours} \\
    \cmidrule(lr){4-7} \cmidrule(lr){8-11} \cmidrule(lr){12-15} \cmidrule(lr){16-19}
    \rowcolor{black!5}
    circuit & $n$ & gates & chip & area & steps & volume & chip & area & steps & volume & chip & area & steps & volume & \cellcolor{tlamber!14}chip & \cellcolor{tlamber!14}area & \cellcolor{tlamber!14}steps & \cellcolor{tlamber!14}volume \\
    \midrule
    \rowcolor{tlplum!4}
    clifford-16q-cx75 & 16 & 39 & $11{\times}11$ & 121 & \textbf{14} & 1,694 & $9{\times}9$ & 81 & \textbf{14} & 1,134 & $7{\times}8$ & 56 & \textbf{14} & \textbf{784} & \cellcolor{tlamber!13}$7{\times}7$ & \cellcolor{tlamber!13}\textbf{49} & \cellcolor{tlamber!13}17 & \cellcolor{tlamber!13}833 \\
    barenco-tof-5 & 9 & 156 & $9{\times}9$ & 81 & \textbf{110} & 8,910 & $9{\times}9$ & 81 & \textbf{110} & 8,910 & $3{\times}15$ & 45 & \textbf{110} & 4,950 & \cellcolor{tlamber!8}$2{\times}15$ & \cellcolor{tlamber!8}\textbf{30} & \cellcolor{tlamber!8}111 & \cellcolor{tlamber!8}\textbf{3,330} \\
    \rowcolor{tlplum!4}
    rc-adder-6 & 14 & 177 & $11{\times}11$ & 121 & \textbf{107} & 12,947 & $11{\times}11$ & 121 & \textbf{107} & 12,947 & $7{\times}10$ & 70 & \textbf{107} & 7,490 & \cellcolor{tlamber!13}$5{\times}10$ & \cellcolor{tlamber!13}\textbf{50} & \cellcolor{tlamber!13}\textbf{107} & \cellcolor{tlamber!13}\textbf{5,350} \\
    qaoa-3reg-24 & 24 & 264 & $13{\times}13$ & 169 & \textbf{74} & 12,506 & $13{\times}13$ & 169 & \textbf{74} & 12,506 & $9{\times}12$ & 108 & \textbf{74} & 7,992 & \cellcolor{tlamber!8}$7{\times}12$ & \cellcolor{tlamber!8}\textbf{84} & \cellcolor{tlamber!8}\textbf{74} & \cellcolor{tlamber!8}\textbf{6,216} \\
    \rowcolor{tlplum!4}
    qaoa-er-16 & 16 & 332 & $11{\times}11$ & 121 & \textbf{89} & 10,769 & $11{\times}11$ & 121 & \textbf{89} & 10,769 & $9{\times}10$ & 90 & 90 & 8,100 & \cellcolor{tlamber!13}$7{\times}9$ & \cellcolor{tlamber!13}\textbf{63} & \cellcolor{tlamber!13}109 & \cellcolor{tlamber!13}\textbf{6,867} \\
    ftcb-adder-28q & 28 & 363 & $15{\times}15$ & 225 & \textbf{161} & 36,225 & $15{\times}15$ & 225 & \textbf{161} & 36,225 & $7{\times}16$ & 112 & 176 & 19,712 & \cellcolor{tlamber!8}$5{\times}16$ & \cellcolor{tlamber!8}\textbf{80} & \cellcolor{tlamber!8}209 & \cellcolor{tlamber!8}\textbf{16,720} \\
    \rowcolor{tlplum!4}
    qcla-mod-7 & 26 & 779 & $15{\times}15$ & 225 & \textbf{200} & 45,000 & $15{\times}15$ & 225 & \textbf{200} & 45,000 & $11{\times}12$ & 132 & 201 & 26,532 & \cellcolor{tlamber!13}$8{\times}12$ & \cellcolor{tlamber!13}\textbf{96} & \cellcolor{tlamber!13}206 & \cellcolor{tlamber!13}\textbf{19,776} \\
    gf2e8-mult & 24 & 853 & $13{\times}13$ & 169 & \textbf{248} & 41,912 & $13{\times}13$ & 169 & \textbf{248} & 41,912 & $7{\times}14$ & 98 & 292 & 28,616 & \cellcolor{tlamber!8}$5{\times}14$ & \cellcolor{tlamber!8}\textbf{70} & \cellcolor{tlamber!8}334 & \cellcolor{tlamber!8}\textbf{23,380} \\
    \rowcolor{tlplum!4}
    adder-8 & 24 & 1,012 & $13{\times}13$ & 169 & \textbf{259} & 43,771 & $13{\times}13$ & 169 & \textbf{259} & 43,771 & $9{\times}12$ & 108 & \textbf{259} & 27,972 & \cellcolor{tlamber!13}$7{\times}12$ & \cellcolor{tlamber!13}\textbf{84} & \cellcolor{tlamber!13}267 & \cellcolor{tlamber!13}\textbf{22,428} \\
    qft-16 & 16 & 1,311 & $11{\times}11$ & 121 & \textbf{282} & 34,122 & $11{\times}11$ & 121 & \textbf{282} & 34,122 & $9{\times}10$ & 90 & \textbf{282} & 25,380 & \cellcolor{tlamber!8}$7{\times}9$ & \cellcolor{tlamber!8}\textbf{63} & \cellcolor{tlamber!8}340 & \cellcolor{tlamber!8}\textbf{21,420} \\
    \midrule
    \rowcolor{tlplum!12}
    Geomean ratio & & & -- & 1.00 & \textbf{1.00} & 1.00 & -- & 0.96 & \textbf{1.00} & 0.96 & -- & 0.60 & 1.03 & 0.61 & \cellcolor{tlamber!20}-- & \cellcolor{tlamber!20}\textbf{0.44} & \cellcolor{tlamber!20}1.13 & \cellcolor{tlamber!20}\textbf{0.49} \\
    \bottomrule
  \end{tabularx}
\end{table*}

As the motivating example in \cref{sec:example} illustrates, volume reductions over \dascot{} could also come from removing unused lanes or adopting a compact static layout.

To isolate the gains from on-demand layouts and the dynamic architecture, \cref{tab:architecture} compares the four configurations introduced in \cref{sec:example}:
\circled[tlplum]{\texttt{A}} Static (DasCot),
\circled[tlplum]{\texttt{B}} Static (DasCot Compacted),
\circled[tlplum]{\texttt{C}} Static (on-demand), and
\circled[tlamber]{\texttt{D}} Ours.
We select \archCircuitCount{} representative benchmarks from different families and report their chip dimensions, area, execution steps, and volume.
Bold marks the lowest area, step count, and volume in each row, including ties.

These results confirm the benefits of two mechanisms introduced by our method:
\begin{itemize}[nosep,leftmargin=*]
  \item \textbf{On-demand layouts.}
  Comparing \circled[tlplum]{\texttt{A}} with \circled[tlplum]{\texttt{B}} shows that only one program benefits from removing unused lanes, whereas comparing \circled[tlplum]{\texttt{A}} with \circled[tlplum]{\texttt{C}} shows that all benchmarks benefit from an on-demand layout.
  Post-processing a conservative layout cannot recover the same savings because lanes rarely remain unused throughout the entire execution.
  \item \textbf{Dynamic layouts.}
  Compared with \circled[tlplum]{\texttt{C}}, \circled[tlamber]{\texttt{D}} uses less area for all of the \archCircuitCount{} benchmarks. Since the only difference here is that \circled[tlplum]{\texttt{C}} uses a single stage while \circled[tlamber]{\texttt{D}} allows different stages to have different layouts, this shows how the dynamic architecture enables more aggressive lane reuse within stages.
  Although our method has the highest step count overall, its area savings outweigh the additional steps, yielding the lowest space--time volume overall.
\end{itemize}

\subsection{Compile Time Comparison}
\label{sec:eval-time}

\begin{figure}[t]
  \centering
  \includegraphics[width=\columnwidth,trim=0 7.5bp 0 3.1bp,clip]{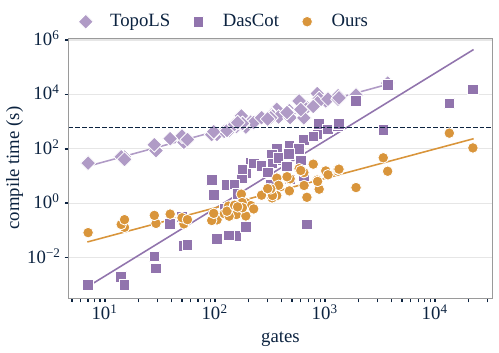}
  \caption{Compile time versus gate count with power-law fits (a), and geometric-mean time (b).}
  \Description{Left: a log-log scatter of compile time against gate count for TopoLS, DasCot, and the solver, each with a fitted line. Right: three bars show geometric-mean compilation times on the same 30 circuits, including 0 stress circuits. Both panels share the logarithmic time axis.}
  \label{fig:time}
\end{figure}

In addition to reducing space--time volume, our method is, to the best of our knowledge, the fastest lattice-surgery compiler.
As discussed in \cref{sec:intro}, fast compilation provides timely feedback to guide circuit synthesis.

\Cref{fig:time} compares compilation time on application circuits that \dascot{} completes within one hour.
We exclude very small circuits because their timings are more sensitive to measurement noise.
The compilers run on the same machines using the experimental setup
in \cref{sec:eval-setup}.
Our method has substantially lower geometric-mean compilation time than either \dascot{} or TopoLS.
The fitted trends also show more gradual growth with gate count for our method than for \dascot{} over the evaluated range.

These speedups mainly stem from the heuristics in \cref{sec:method}, which limit search effort at each stage.
Whereas TopoLS uses Monte Carlo tree search for placement, we use simulated annealing for faster convergence with a controllable search budget.
Whereas \dascot{} uses breadth-first search to find shortest routing paths, we assign one deterministic route to each interaction.
We also select the floorplan using a proxy model instead of grid search.
Together, these choices make our method $\headlineDascotSpeedup\times$ faster than \dascot{} and $\headlineTopolsSpeedup\times$ faster than TopoLS in geometric mean.

The implementation also exploits graph sparsity: readiness updates
follow dependency edges, and placement moves rescore only the
layers they affect.
Further implementation details are available in our open-source
tool.

\subsection{Breakdown and Utilization Analysis}
\label{sec:eval-breakdown}

\begin{figure}[t]
  \centering
  \begin{tikzpicture}
    \node[anchor=south west,inner sep=0] (plot) at (0,0)
      {\includegraphics[width=\columnwidth]{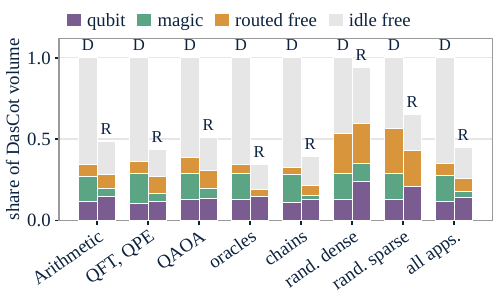}};
    \begin{scope}[x={(plot.south east)},y={(plot.north west)}]
      \fill[white] (.8197,.0660) rectangle (.8574,.1681);
      \fill[white] (.8603,.1104) rectangle (.9209,.2382);
      \fill[white] (.9164,.1764) rectangle (.9291,.2527);
      \node[anchor=base east,rotate=35,inner sep=0,text=tlnavy,
        font=\fontsize{9}{10}\selectfont] at (.923,.200) {Geomean};
      \node[rotate=90,inner sep=0,text=tlnavy,font=\fontsize{8}{9}\selectfont]
        at (.8925,.631) {\pgfmathprintnumber[fixed,precision=1]{\headlineDascotUtilization}\%};
      \node[rotate=90,inner sep=0,text=tlnavy,font=\fontsize{8}{9}\selectfont]
        at (.930,.457) {\pgfmathprintnumber[fixed,precision=1]{\headlineRVLaneUtilization}\%};
    \end{scope}
    \pgfresetboundingbox
    \path[use as bounding box] ([yshift=3.5bp]plot.south west)
      rectangle ([yshift=-5.5bp]plot.north east);
  \end{tikzpicture}
  \caption{Volume breakdown for \dascot (D) and our method (R). Bars average tile-type contributions normalized to each circuit's \dascot volume, grouped by family.}
  \Description{Stacked bars per benchmark family split the volume of DasCot and the solver into qubit, magic, routed, and idle tile-steps.}
  \label{fig:breakdown}
\end{figure}

To understand the sources of volume reduction, \cref{fig:breakdown} separates space--time volume into data-qubit tiles (\qsq{}), magic-state tiles (\msq{}), routed free tiles (\psq{}), and idle free tiles (\fsq{}), relative to \dascot{}.
Our method reduces the idle routing-space contribution by sharing lanes
and adjusting physical spacing.
Within each program, routing-space utilization is the routed-free contribution (\psq{}) divided by the sum of routed-free and idle-free contributions (\psq{} + \fsq{}).
Geometric-mean utilization increases from \headlineDascotUtilization\% for \dascot{} to \headlineRVLaneUtilization\% for our method, an improvement of \headlineUtilizationImprovement{}\%.

The routed-free contribution increases in most families even as total volume decreases.
This illustrates why minimizing path length alone is insufficient:
a longer route can be worthwhile when it allows interactions to use
a smaller shared layout.
By accounting for both concurrency and the coordinates retained after
compaction, the router can make this tradeoff rather than optimizing
wire length in isolation.

\subsection{Ablation Study}
\label{sec:eval-baseline}
\label{sec:eval-ablation}

\begin{figure}[t]
  \centering
  \includegraphics[width=\columnwidth,trim=0 4.1bp 0 1.9bp,clip]{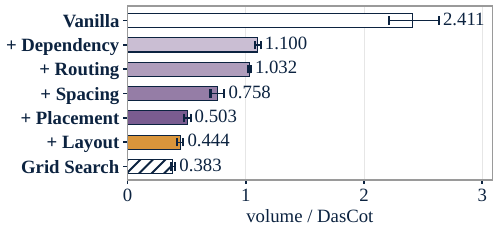}
  \caption{Cumulative ablation on 49 applications: geometric-mean volume relative to \dascot, with standard errors. Grid Search covers the 48 applications whose exhaustive search has finished.}
  \Description{Vanilla: serial baseline. + Dependency: add dependency-aware scheduling. + Routing: add criticality-driven routing on the fixed ASAP mapping. + Spacing: add on-demand spacing. + Placement: add placement on the square paired-lane array. + Layout: add demand-based layout selection. Grid Search: exhaustive shape and lane-budget selection.}
  \label{fig:ablation-naive}
\end{figure}

The preceding comparisons show the strengths of our method and its dynamic architecture.
Realizing these benefits requires a specialized compiler.
\Cref{fig:ablation-naive} examines each compiler component's contribution through a cumulative ablation that adds the methods introduced in \cref{sec:method}:
\begin{itemize}
  \item \textbf{Vanilla}: Execute one gate per step.

  \item \textbf{+ Dependency}: Add dependency-aware scheduling
  (\cref{sec:form-execution}).

  \item \textbf{+ Routing}: Add criticality-driven routing
  (\cref{sec:routing}) on the fixed grid, without compaction or walks.

  \item \textbf{+ Spacing}: Add on-demand spacing
  (\cref{sec:dynamic-mapping}).

  \item \textbf{+ Placement}: Add placement on a square Standard
  layout (\cref{sec:mapping}).

  \item \textbf{+ Layout}: Add layout selection (\cref{sec:layout}),
  yielding our method.

  \item \textbf{Grid Search}: Exhaustively select the shape and lane budgets
  with the placement and routing algorithms held fixed.
\end{itemize}

Each addition from \textbf{+ Dependency} through \textbf{+ Layout}
retains the preceding components.
The successive reductions show the benefits of scheduling concurrent
gates, placing qubits to reduce contention, and matching the
floorplan to circuit demand.

Among these additions, routing alone provides a smaller improvement on the fixed grid.
Although route selection accounts for retained coordinates, reducing
their number cannot shrink the charged area until compaction is
enabled.
Adding on-demand spacing converts this reduction in retained space
into a smaller physical layout.
This illustrates the coupling between routing and compaction:
routing determines which coordinates are needed, and compaction
turns that choice into area savings.

Exhaustive configuration selection explores further volume reductions
by repeatedly invoking placement and routing across candidate shapes
and lane budgets.
Predicting a single configuration avoids these repeated compilations,
at the cost of potentially missing a better configuration.

\subsection{Sensitivity Study}
\label{sec:eval-sensitivity-main}
Although effective, heuristics often require hyperparameter tuning.
Tuning each benchmark separately or repeatedly trying more random seeds can erode the advantage over exhaustive search.
To assess robustness, we study sensitivity to the following parameters:
\begin{itemize}[nosep,leftmargin=*]
  \item \textbf{Collision weight $w_\chi$} (\cref{sec:mapping}): penalty for estimated route overlaps.
  \item \textbf{Lane penalty $\lambda$} (\cref{sec:routing}): cost of additionally retained coordinates.
  \item \textbf{Placement effort $\eta$} (\cref{sec:mapping}): a cap of $\eta n$ annealing proposals.
  \item \textbf{Lane budgets $\Delta_H,\Delta_W$} (\cref{sec:layout}): extra retained rows and columns, varied together.
\end{itemize}

We vary each parameter in turn, sweeping the two lane budgets together and keeping the other parameters at their default settings.
Each experiment recompiles the complete program, including compaction
and modeled walks.

\begin{figure*}[!t]
  \centering
  \includegraphics[width=\textwidth,trim=0 5.5bp 0 7bp,clip]{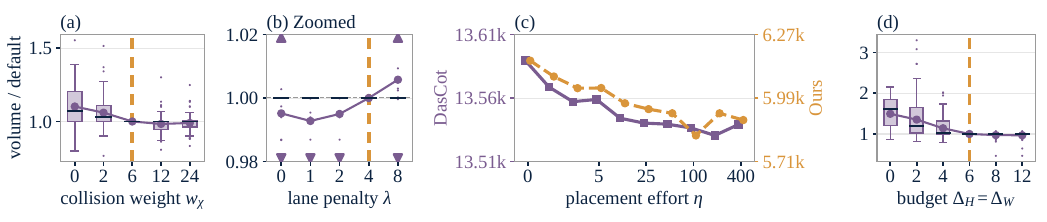}
  \caption{Sensitivity to collision weight (a), lane penalty (b), placement effort (c), and lane budgets (d). Boxes and circles in (a,b,d) show volume distributions and geometric means relative to the default configuration of our method; triangles mark off-scale values in (b).}
  \Description{Four panels in one row show collision-weight, lane-penalty, placement-proposal, and lane-budget sweeps. The lane-penalty axis is zoomed around parity, with triangles marking off-scale observations. The placement panel uses purple DasCot and orange our method curves with separate volume axes. The final panel varies the additional-row and additional-column budgets together.}
  \label{fig:sens-collision-lane}
\end{figure*}

In the other experiments, we use $w_\chi=\text{\sensitivityCollisionDefault}$, $\lambda=\text{\sensitivityLaneDefault}$, $\eta=\text{\sensitivityPlacementDefault}$, and $\Delta_H=\Delta_W=\text{\sensitivityBudgetDefault}$.
Across the settings tested in \cref{fig:sens-collision-lane}, the maximum absolute percentage change in geometric-mean volume relative to the default configuration is \sensitivityVolumeDeviation\%.

\smallskip\noindent\textbf{Collision Weight.}
Removing the placement collision term increases geometric-mean volume
in \cref{fig:sens-collision-lane}a.
Increasing its weight beyond the default provides a smaller additional
benefit.
Accounting for contention therefore matters more than fine-tuning
the precise weight.

\smallskip\noindent\textbf{Lane Penalty.}
Geometric-mean volume changes little across the evaluated penalty
settings, including when the penalty is disabled.
The zoomed axis in \cref{fig:sens-collision-lane}b makes these small
aggregate changes visible.
Individual circuits can be more sensitive, as indicated by the
markers at the axis boundaries.

\smallskip\noindent\textbf{Placement Proposals.}
\Cref{fig:sens-collision-lane}c shows that refining the spectral
initialization improves volume, but additional search beyond the
default effort does not consistently improve the compiled program.
Annealing optimizes a placement proxy rather than final routed
volume, so more proposals do not guarantee a better result.
The figure also evaluates \dascot{}'s annealer under the same
proposal caps and on the same inputs, with all declared qubits
placed and no time limit.

\section{Case Study: Feedback to Circuit Synthesis}
\label{sec:casestudy}
\begin{figure}[!t]
  \centering
  \includegraphics[width=\columnwidth,trim=0 4bp 0 5.3bp,clip]{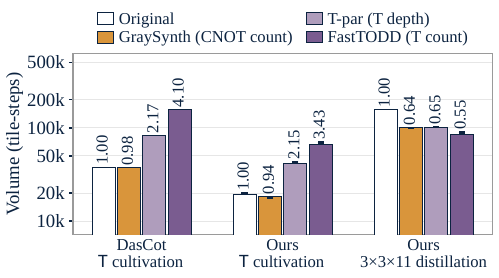}
  \caption{Volume of the 21-qubit $\mathrm{GF}(2^7)$ multiplier under circuit optimizations (log scale): \dascot{} and our method with \texttt{T} cultivation, and our method with $3\times3\times11$ distillation. Numbers are ratios to the original circuit.}
  \Description{Grouped volume bars for DasCot and two our method configurations. White denotes the original circuit, orange denotes GraySynth, light purple denotes T-par, and dark purple denotes FastTODD.}
  \label{fig:optvariants}
\end{figure}

The preceding evaluation shows our method's effectiveness and efficiency as a lattice-surgery compiler.
As discussed in \cref{sec:intro}, these compilers also provide feedback and technology-dependent insights to guide circuit synthesis.

\begingroup\tolerance=100\emergencystretch=1em
We use $\mathrm{GF}(2^k)$ multipliers to compare three classes of circuit optimizers and study their implications for lattice surgery on superconducting-qubit platforms.
GraySynth~\cite{Amy_2018} targets \CX{} count, T-par~\cite{amy2014polynomial} targets \texttt{T} depth, and FastTODD~\cite{vandaele2024lower} targets \texttt{T} count.
We compare the feedback from our method with that from \dascot{}.
\par\endgroup

\Cref{fig:optvariants} shows the results for the 21-qubit multiplier.
Both compilers account for tradeoffs between \texttt{T} and \CX{} gates and show similar trends when magic-state supply is unconstrained, our model for the \texttt{T}-cultivation setting~\cite{gidney2024magic}.
In this regime, aggressive \texttt{T}-count optimization introduces substantial ancilla-qubit and \CX{} overhead, so FastTODD consumes more volume than GraySynth, which targets only \CX{} count.

An advantage of our method and its dynamic scheduling is that we can readily evaluate different assumptions about \texttt{T}-state supply.
For example, modeling the $3\times3\times11$ distillation factories from LaSsynth~\cite{tan2024sat} requires only adding dependencies to the TDG based on their throughput.
Under this supply constraint, \texttt{T}-count optimization becomes beneficial and outperforms optimization of \CX{} count alone.

Future work could use our method as a reward model to guide synthesis at a finer granularity and search for better tradeoffs.
Its efficiency and accuracy make it well suited to this role.

\section{Related Work}
\label{sec:related}

\begingroup\emergencystretch=0.5em
LibLSQECC~\cite{watkins2024high, leblond2023realistic, leblond2025comparison}, \dascot~\cite{dascot2024}, Ecmas+~\cite{zhu2025ecmas}, and resource-allocating compilers~\cite{robertson2025resourceallocating} map and schedule lattice surgery.
TISCC generates physical circuits for surface-code patch operations on trapped-ion processors~\cite{leblond2023tiscc}.
SAT-based synthesis and spacetime-routing compilers optimize three-dimensional realizations~\cite{tan2024sat, liao2026koval, topologiq2024, topols2025, hamada2024efficient, hamada2026bounded}.
SpiderLS applies full ZX reduction and lowers the resulting interactions to Pauli-product measurements before routing them with fixed data-patch positions~\cite{kim2026spiderls}.
TQEC lowers three-dimensional block graphs to physical Stim circuits~\cite{suau2026tqec}, and Hao et al. integrate physical-circuit compilation and simulation to study algorithmic break-even~\cite{hao2025breakeven}.
Patchwork constructs physical measurement circuits on a prescribed layout with narrow corridors and partitions the merged patches in spacetime for parallel decoding~\cite{chamberland2026patchwork}.
Herzog et al. extend mapping and routing to other code substrates~\cite{herzog2025beyond} and develop pipe diagrams for the color code~\cite{herzog2026color}.
Analytical resource estimators instead provide cost estimates without constructing routes~\cite{harrigan2024expressing, beverland2022assessing, huggins2025flasq}.
\par\endgroup

\begingroup\emergencystretch=0.5em
Layout design trades data density against space for concurrent routes.
Litinski's compact blocks~\cite{litinski2019game}, checkerboard placement~\cite{lao2019mapping}, O3LS's squeezed layout~\cite{zhu2026o3ls}, SPARO's expandable routing area~\cite{kan2025sparo}, distillation-adaptive layouts~\cite{sharma2025spacetime}, LAPBC~\cite{hirano2025locality}, and Q-Spellbook~\cite{chatterjee2025qspellbook} explore different arrangements of data and routing tiles.
Unlike O3LS's Pauli-product scheduling, our main flow's compaction preserves a routed \CX{}/\texttt{T} schedule and the endpoint orientations selected during placement.
The extension in \cref{sec:appendix-pbc} applies stage compaction to multi-qubit Pauli operations.
Dense storage~\cite{kobori2025lsqca, gidney2025yoked, low2026denser}, workload-specific floorplans~\cite{aghosh2026workload}, and multiprogramming~\cite{aghosh2026notile} address how chip capacity is partitioned and reused.
Braiding and its connection to lattice surgery~\cite{javadiabhari2017optimized, hua2021autobraid, paler2025untangling}, distillation~\cite{ding2018magic, bravyi2012distillation, jones2013multilevel, litinski2019magic, gidney2019efficient}, and circuit synthesis and instruction scheduling~\cite{wang2024optimizing, bharadwaj2026cphase, kuo2026ftqcs} optimize the operations performed on the chip.
Alternative connectivity~\cite{ueno2024bypass, litinski2022active}, temporal protocols~\cite{fowler2012timeoptimal, chamberland2022twistfree}, and fixed-layout scheduling~\cite{silva2024multiqubit} address complementary resource costs.
Beverland et al. study edge-disjoint \CX{} routing on shared space~\cite{beverland2022edp}.
\par\endgroup

Related approaches adapt patch allocation and placement during execution.
CircLS allocates each patch from first to last use~\cite{zhang2026circls}, while RushHour reshapes a staircase corridor with patch slides to unblock ready gates~\cite{tornow2026rushhour}.
Herzog et al. move logical qubits by teleportation during a \CX{}, with a proof-of-concept compiler for the color code~\cite{herzog2025movablelogical}.
PureMagic dynamically shares ancilla patches between routing and magic-state cultivation~\cite{puremagic2026, gidney2024magic}.
FlowRouter combines maximum-flow routing of Clifford layers in three-dimensional spacetime with runtime cultivation on idle patches and delay insertion when states are unavailable~\cite{iacobacci2026flowrouter}.
Walking surface-code constructions~\cite{mcewen2023relaxing} and dynamic-code memory experiments~\cite{eickbusch2025dynamic} motivate patch movement.
Our method forms stages during routing and compacts them, preserving gate grouping, all data qubits, and their order while changing spacing through modeled bulk walks.
Classical layout compaction also preserves geometric constraints~\cite{Liao1983Compaction}, while two-layer channel routing permits wires to cross on separate layers~\cite{Yoshimura1982Channel}.
Our tile-disjoint paths cannot cross within a step, so area savings may incur serialization.

\section{Conclusion}
\label{sec:conclusion}

Our method combines a dynamic architecture with compaction-aware routing and demand-driven floorplanning.
Across our application circuits, it reduces geometric-mean space-time volume by \headlineVolumeSaving\% and compiles $\headlineDascotSpeedup\times$ faster than \dascot.

\section*{Acknowledgments}
The authors thank Dr. Daniel Tan for valuable discussion. H. Wang and J. Cong were supported by the National Science Foundation under Award Nos.~2533041 and OMA-2016245, the industrial partners of the Center for Domain-Specific Computing (\url{https://cdsc.ucla.edu/partners/}), and a gift from Google LLC. 
Z. Chen and Z. Zhang acknowledge support from the National Science Foundation under Award No.~2529338.
J. Cong has financial interest in Google and IBM.

\bibliographystyle{acm-shortauthors}
\bibliography{reference}

\clearpage
\appendix
\setcounter{dbltopnumber}{2}\renewcommand{\dbltopfraction}{0.85}
\section{Local Resynthesis with LaSsynth}
\label{sec:appendix-resynth}

The compiler in \cref{sec:method} optimizes routing space but realizes gates individually.
Synthesizing a group of gates together may yield a shallower implementation.
In this section, we introduce local resynthesis in two steps.
We first extract small windows from the tile program for independent optimization.
We then use LaSsynth~\cite{tan2024sat} to search for the shallowest pipe diagram for each window's Clifford computation.
Fixed footprints and spatial ports let the search focus on depth.

\begin{figure*}[!t]
  \centering
  \includegraphics[width=\textwidth]{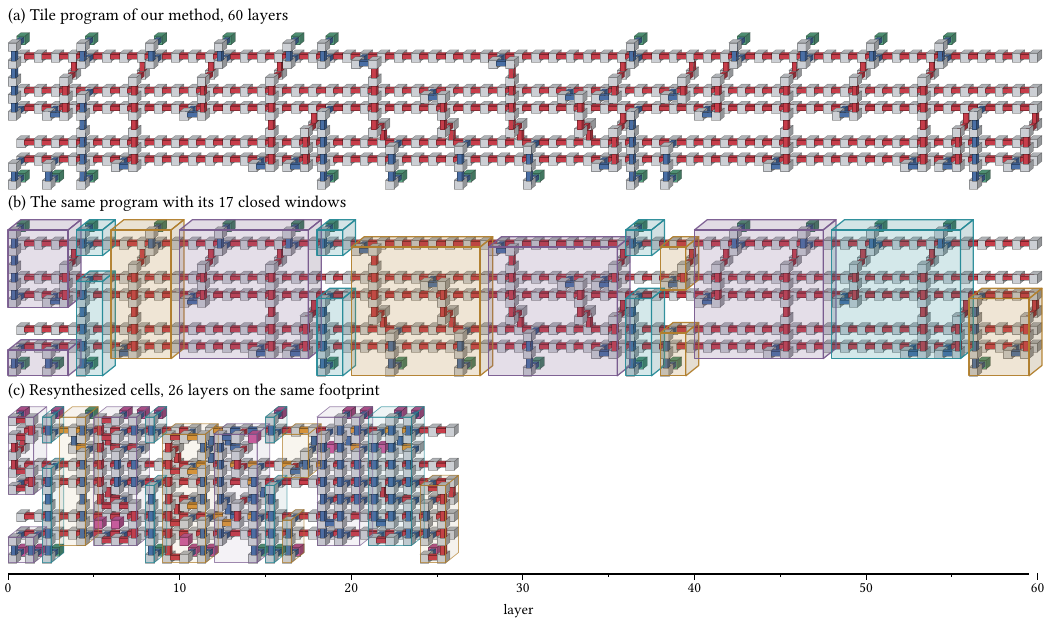}
  \caption{Local resynthesis of tof-3 on a common time scale: (a) the original tile program; (b) closed windows with fixed spatial ports, shaded by slab; (c) resynthesized cells with reduced depth and the same footprint. Red and blue pipes indicate the two orientations, grey cubes patches, and green cubes magic states. In (c), amber pipes indicate Hadamard walls and pink cubes Y cubes; their additional implementation costs are omitted from the volume estimate.}
  \label{fig:resynth-windows}
  \Description{Three rows of a 3D pipe diagram on one time scale with a layer ruler. The top row is the original tile program over 60 layers, the middle row shades 17 window boxes over the same program, and the bottom row shows the resynthesized cells over 26 layers, less than half as long.}
\end{figure*}

\smallskip\noindent\textbf{Window Extraction.}
Ideally, we would apply exact SAT-based synthesis to the entire program, allowing optimizations to span all gate boundaries.
However, the cost of exact synthesis makes this impractical at scale, so we restrict the search to small windows.
For these experiments, we limit each window to 16 tiles, 4 steps, and 12 ports.
These bounds keep each search small, at the cost of excluding optimizations that require a larger region.

Each window spans a rectangular tile region and contiguous steps within one fixed-mapping stage of \cref{sec:form-lanes}.
We require the window to be closed: every route in that interval must lie entirely inside or outside the region.
This condition lets us change the routes inside a window without disturbing routes outside it.
Windows spanning the same steps form a slab and share no qubits, so their synthesized implementations can execute in parallel.
A dynamic program chooses the slab boundaries, as illustrated for tof-3 in \cref{fig:resynth-windows}b.

\smallskip\noindent\textbf{Window Resynthesis.}
We pass each extracted window to LaSsynth with fixed spatial port positions, allowing a replacement to reconnect to the surrounding program.
For windows containing \texttt{T} gates, we represent each \texttt{T} merge as a magic-state input port, leaving only the Clifford skeleton for LaSsynth to resynthesize.
The solver tests increasing depths starting at one layer, with a limit of 60\,s per call.
If it finds no shallower solution, we retain the original pipes, ensuring that resynthesis never increases the input program's depth.

The deepest resynthesized window determines each slab's estimated depth (\cref{fig:resynth-windows}c).
With the footprint fixed, fewer layers reduce the volume objective in \cref{sec:form-objective}.
Note that LaSsynth may introduce Hadamard walls and Y cubes, whose additional implementation costs are omitted from this estimate.

\begin{table*}[!t]
  \caption{Volume, in patches times $d$ rounds, and single-core compile time of local resynthesis on Clifford+\texttt{T} circuits with magic states on demand, against the runs of \cref{sec:eval}. \dascot and our method count twice the chip area times steps, TopoLS its cube bounding box, and bold marks the minimum in each row.}
  \label{tab:resynth}
  \centering\small
  \setlength{\tabcolsep}{8.85pt}  %
  \begin{tabular}{lrr|rrrr|rrrr}
    \toprule
    \rowcolor{tlplum!15}
    & & & \multicolumn{4}{c|}{Volume} & \multicolumn{4}{c}{Compile time (s)} \\
    \rowcolor{tlplum!15}
    circuit & $n$ & gates & \dascot & TopoLS & Ours & + LaSsynth & \dascot & TopoLS & Ours & + LaSsynth \\
    \midrule
    barenco-tof-3 & 5 & 52 & 6,156 & 2,070 & 1,440 & \textbf{612} & \textbf{0.03} & 169 & 0.17 & 616 \\
    mod5-4 & 5 & 56 & 7,128 & 2,160 & 1,840 & \textbf{800} & \textbf{0.03} & 198 & 0.19 & 698 \\
    ftcb-qft-4q & 4 & 93 & 5,684 & 1,971 & 1,920 & \textbf{944} & 7.95 & 344 & \textbf{0.20} & 233 \\
    mod-mult-55 & 9 & 97 & 7,614 & 5,670 & 3,120 & \textbf{2,310} & 2.17 & 386 & \textbf{0.84} & 466 \\
    barenco-tof-4 & 7 & 104 & 11,988 & 4,320 & 3,744 & \textbf{2,616} & \textbf{0.05} & 363 & 0.71 & 367 \\
    vbe-adder-3 & 10 & 128 & 17,424 & 5,796 & 5,616 & \textbf{3,600} & 5.31 & 496 & \textbf{1.17} & 423 \\
    qft-4 & 5 & 134 & 17,010 & 5,175 & 4,280 & \textbf{2,200} & \textbf{0.07} & 460 & 0.41 & 460 \\
    barenco-tof-5 & 9 & 156 & 17,820 & 8,757 & 6,496 & \textbf{3,724} & \textbf{0.06} & 618 & 1.31 & 488 \\
    rc-adder-6 & 14 & 177 & 25,894 & 10,854 & 11,520 & \textbf{8,360} & 19.5 & 831 & \textbf{2.35} & 438 \\
    csla-mux-3 & 15 & 191 & 16,456 & 10,692 & 7,200 & \textbf{5,350} & 13.0 & 982 & \textbf{3.04} & 446 \\
    gf2e4-mult & 12 & 211 & 24,926 & 11,277 & 10,710 & \textbf{8,370} & 31.3 & 882 & \textbf{2.27} & 529 \\
    \midrule
    \rowcolor{tlplum!4}
    geomean & & & 12,606 & 5,194 & 4,253 & \textbf{2,559} & 0.82 & 453 & \textbf{0.75} & 454 \\
    \bottomrule
  \end{tabular}
\end{table*}

Local resynthesis cuts our method's volume by 40\% in geometric mean, placing every tabulated circuit below both baselines (\cref{tab:resynth}).
The benefit extends to nine Clifford-only circuits, whose geometric-mean volume falls from 710 to 459, also below both baselines.
The gains depend on window coverage: fully covered circuits improve most, while wide steps in gf2e4-mult exceed the budget and remain unchanged.
Longer searches cannot help these excluded steps, and no synthesis call times out.
These savings require minutes of synthesis per circuit, compared with seconds for constructing the input tile program.
We cache synthesis results for repeated windows, so runtime depends on the number of distinct windows and the cost of ruling out shallow depths.
These searches could run in parallel, although measured sequentially here.

\section{Hierarchical Compilation}
\label{sec:appendix-hierarchical}

The compiler in \cref{sec:method} operates on a flattened gate list, repeating placement and routing work across calls to the same subroutine.
Hierarchical representations, such as Qualtran bloqs~\cite{harrigan2024expressing}, expose these repetitions and enable compilation reuse.
Robertson et al.~\cite{robertson2025resourceallocating} recursively reuse compiled lattice-surgery objects as gates in larger compilations, but their objects have fixed dimensions and input/output interfaces.
In our dynamic architecture, however, physical spacing must adapt to the routing demand of overlapping calls.
Reusing fixed physical layouts would therefore limit the flexibility of stage-level compaction.

To enable hierarchical reuse without sacrificing dynamic routing-space allocation, we separate cell compilation from physical compaction.
We first construct a shared virtual floorplan that fixes qubit assignments and preserves row and column orders across cells.
Each distinct cell is then routed once on this floorplan, and parent schedules are composed bottom-up from cached implementations, resolving routing conflicts as needed.
Only at the root do we form stages and compact them according to their combined routing demand.
Walks connect successive stages, allowing physical spacing to vary while reusing the compiled cells.
This reuse could also amortize local resynthesis, as discussed in \cref{sec:appendix-resynth}.

\begin{figure*}[!t]
  \centering
  \includegraphics{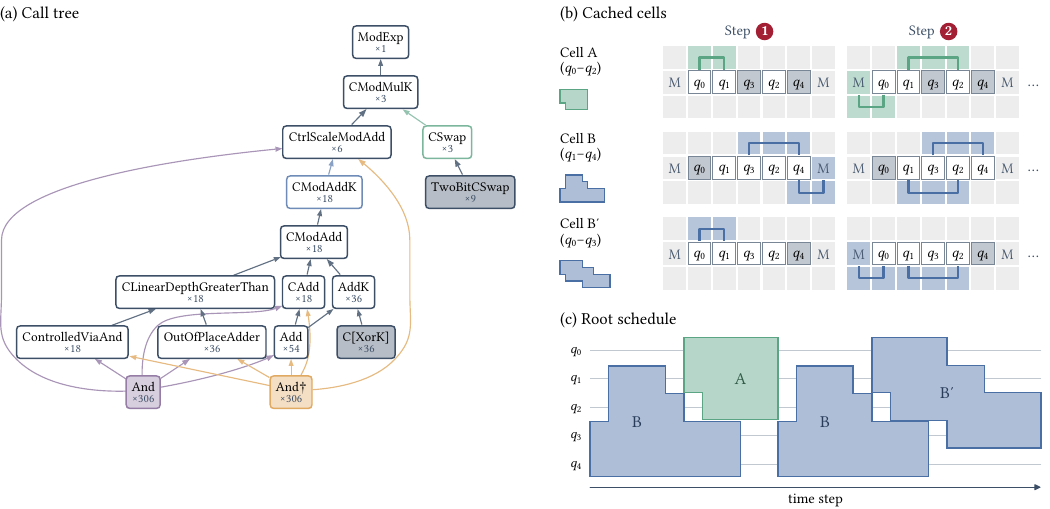}
  \caption{Bottom-up compilation of modexp-3: (a) the call tree, compiled from children to parents, with call counts; (b) cells routed once on the shared floorplan, where B$'$ has the logic of B on other qubits and is cached separately; (c) the root schedule, where B is reused and routing conflicts delay calls. Panels (b) and (c) are schematic, with colors marking cell logic.}
  \label{fig:hierarchy}
  \Description{Left, the modexp-3 call graph with upward arrows from compiled children to their parents, ending at ModExp; filled boxes are leaves and outlined boxes are parents. Top right, cells A, B, and B-prime, each shown as a block of qubits over steps beside its first routed steps on a shared five-qubit floorplan; B-prime has the logic of B on other qubits, with different routes and a different block shape. Bottom right, a root schedule built from these blocks: B, A, B again, and B-prime, separated by gaps from routing conflicts.}

\end{figure*}

\smallskip\noindent\textbf{Cells.}
We lower each Qualtran benchmark to \CX{} and \texttt{T} gates while preserving its subroutine call tree.
Each leaf contains primitive gates, while a parent consists of child calls and possibly gates executed directly at that level (direct gate runs).
We cache the routed schedule of each distinct leaf, direct gate run, and parent for reuse across repeated calls.
Since routing depends on qubit placement, two cells can share a cached implementation only if they perform the same operations on the same global qubits.
For example, B and B$'$ in \cref{fig:hierarchy}b implement the same logic on different qubits and therefore require separate implementations.
Reusing a parent additionally requires the same ordered sequence of child calls and direct gate runs.

\smallskip\noindent\textbf{Shared Floorplan.}
Because cached cells include routing paths, reusing them requires compatible qubit placements.
If each cell were placed independently, the same qubits could have different row or column orders across cells.
Such placements cannot be reconciled by the walks in \cref{sec:architecture}, which adjust qubit spacing but preserve their relative order.
We therefore construct a single virtual floorplan for the entire program and compile all cells against this shared placement.
This ensures a consistent spatial arrangement for composing their routed schedules, while allowing physical spacing to vary during stage compaction.

To construct the shared floorplan efficiently, we run the placer of \cref{sec:mapping} once using dependency layers from distinct leaf cells.
Each layer is weighted by its number of invocations in the expanded call tree, preserving the relative frequency of interactions without repeatedly processing identical cells.
By considering gates only at the leaves, we also avoid double-counting operations represented by their ancestors.
However, this approach introduces two limitations.
First, the shared placement restricts cell-level optimization: even qubits unused by a cell remain routing obstacles, as illustrated by cell A in \cref{fig:hierarchy}b.
Second, placement considers each leaf's dependency layers independently and therefore only approximates contention between overlapping calls.

\begin{table}[!t]
  \caption{Hierarchical (Hier.) versus flat compilation of Qualtran circuits: volume ratio (Hier./flat) and single-core compile time in seconds, faster in bold. Calls count all invocations, including the root and gate runs; cells count distinct implementations.}
  \label{tab:hierarchy}
  \centering\small
  \setlength{\tabcolsep}{5pt}
  \begin{tabular*}{\columnwidth}{@{\extracolsep{\fill}}lrrrr|r|rr}
    \toprule
    \rowcolor{tlplum!15}
    & & & & & Volume & \multicolumn{2}{c}{Time (s)} \\
    \rowcolor{tlplum!15}
    circuit & $n$ & gates & calls & cells & ratio & Flat & Hier. \\
    \midrule
    modexp-3 & 22 & 5,040 & 1,372 & 170 & 1.11 & 23.19 & \textbf{3.10} \\
    modmul-4 & 24 & 2,904 & 752 & 162 & 1.05 & 13.74 & \textbf{2.49} \\
    gf2inv-4 & 32 & 775 & 65 & 63 & 1.16 & 3.15 & \textbf{2.52} \\
    modadd-8 & 34 & 531 & 132 & 131 & 1.10 & 2.18 & \textbf{1.80} \\
    add-16 & 47 & 252 & 61 & 61 & 0.99 & 1.20 & \textbf{0.98} \\
    hubbard-2 & 19 & 336 & 91 & 51 & 1.27 & 1.06 & \textbf{0.49} \\
    qrom-6 & 17 & 921 & 239 & 55 & 1.04 & 2.81 & \textbf{0.57} \\
    gf2mul-8 & 24 & 981 & 66 & 66 & 1.34 & 4.69 & \textbf{3.00} \\
    \midrule
    \rowcolor{tlplum!4}
    geomean & & & & & 1.13 & 3.72 & \textbf{1.52} \\
    \bottomrule
  \end{tabular*}
\end{table}

\smallskip\noindent\textbf{Bottom-Up Compilation.}
Given the shared floorplan, we compile the call tree from leaves to root, without a cell-size cutoff (\cref{fig:hierarchy}a).
We first route each distinct leaf or direct gate run once, caching its per-step schedule and routes (\cref{fig:hierarchy}b).
We then construct each parent by composing its cached children and direct runs, and cache the resulting schedule for future reuse.
For example, repeated \texttt{CModAddK} calls reuse both their descendants and composed schedules, as do the two calls of B in \cref{fig:hierarchy}c.

Although cells share a floorplan, their routes may conflict when multiple calls overlap in time.
To compose their schedules, we process ready child steps in call order, respecting dependencies and tile availability while preserving each call's internal step order and gate grouping.
When routes conflict, we first attempt to reroute them using the candidate search of \cref{sec:routing}, followed by a general shortest-path search.
If neither succeeds, the conflicting step is delayed.
A call releases each qubit after its last use, allowing subsequent calls to begin before the earlier call finishes (\cref{fig:hierarchy}c).

Hierarchical compilation achieves a $2.45\times$ geometric-mean speedup at the cost of a 12.7\% increase in space--time volume (\cref{tab:hierarchy}).
The speedup comes from two sources: reusing compiled cells and reducing placement effort.
Reuse provides the largest gains for modexp-3, modmul-4, and qrom-6, which contain many repeated subroutines, and also benefits the smaller hubbard-2 circuit.
Even add-16, where all cells are distinct, compiles faster because leaf-local placement is cheaper than placement on the full dependency graph.

The volume overhead also has two sources.
First, cached schedules preserve the gate groupings within each cell.
During composition, steps can be rerouted or delayed, but gates cannot be regrouped to better accommodate overlapping calls.
Second, leaf-local placement does not capture routing contention between calls, potentially increasing their combined routing demand.
For gf2mul-8, using the flattened circuit's placement reduces the volume overhead from 34\% to 18\%, but brings compilation time close to the flat baseline.
This illustrates the tradeoff between compilation efficiency and global optimization quality.

\begin{figure*}[!t]
  \centering
  \includegraphics{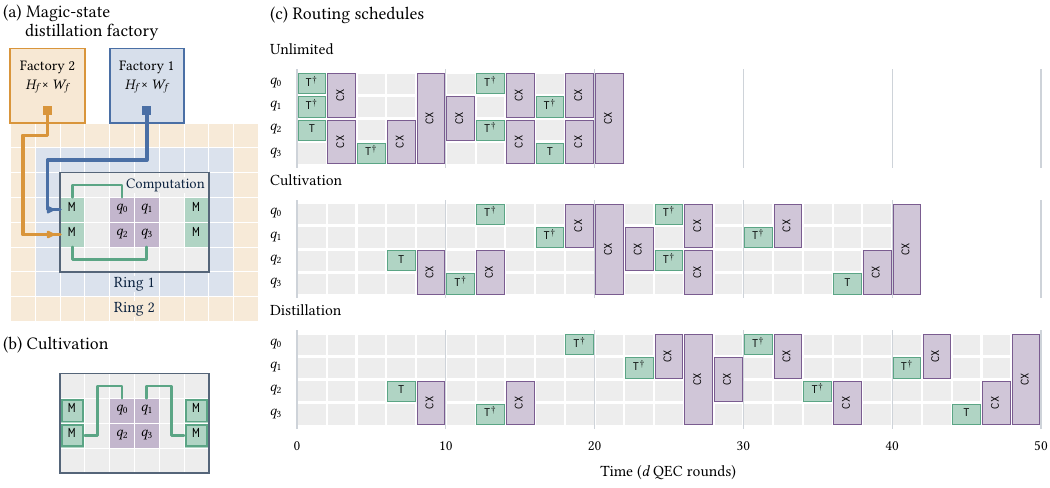}
  \caption{Magic-state supply through (a) distillation factories and (b) in-place cultivation~\cite{gidney2024magic}. (c) Routing ftcb-adder-4q on a shared clock in units of $d$ QEC rounds. Each box spans $2d$ rounds: green for \texttt{T}/\texttt{T}$^\dagger$, purple for \CX{}, and gray for idle qubits. Distillation uses two factories with period $11d$ rounds. Cultivation uses an illustrative stochastic trace.}
  \label{fig:magic-supply}
  \Description{Panels a and b are stacked on the left, with panel c on the right. Panel a shows separate distillation factories with symbolic height H sub f and width W sub f above concentric delivery rings surrounding a four-by-six-tile computation grid. Amber Factory 2 sits on the left and uses outer Ring 2, while blue Factory 1 sits on the right and uses inner Ring 1. Shaded labels sit directly inside the computation grid and both rings. Four data qubits occupy the central sites, with magic-state tiles at both ends of each site row. Disjoint paths deliver to the upper-left and lower-left magic tiles at different steps. Panel b shows in-place cultivation and representative routes from qubits zero and one to the two lower magic tiles. Panel c compares three verified schedules of the complete four-qubit adder benchmark on identical axes from zero to fifty units of d QEC rounds. Each route box represents one routed step of two units, with small white gaps between adjacent step cells. Its normalized gate list has ten CNOT and eight T or T-dagger gates. Green boxes show magic-state routes, and purple boxes labeled CX span the participating qubit rows. Gray idle cells continue on every qubit row until that schedule completes. Unlimited supply completes the circuit in eleven steps or twenty-two time units, the illustrative cultivation trace in twenty-one steps or forty-two units, and shared distillation in twenty-five steps or fifty units. CNOT routes execute between state arrivals, while dependent CNOT routes wait for preceding magic-state gates.}
\end{figure*}

\section{Magic-State Supply}
\label{sec:appendix-magic}

Each \texttt{T}/\texttt{T}$^\dagger$ gate requires a fresh magic state in the routing model of \cref{sec:formulation}.
While the volume comparisons in \cref{sec:eval-full} assume states are available on demand, finite supply can introduce additional resource costs and execution delays.
We consider two supply mechanisms: distillation in external factories and in-place cultivation, illustrated in \cref{fig:magic-supply}.

\smallskip\noindent\textbf{Magic-State Distillation.}
Distillation produces magic states in dedicated factories, requiring additional chip area and delivery routes.
We model $F$ factories outside the $H\times W$ computation grid, each occupying $H_f\times W_f$ tiles and producing one state every $P$ units of $d$ QEC rounds.
Together, the factories occupy $FH_fW_f$ tiles and provide an aggregate production rate of $F/P$ states per time unit.
Since each routed step spans two such units, the cumulative number of states produced by the start of step $t$ is
\[
 N_{\mathrm{supply}}(t)=\left\lfloor\frac{2tF}{P}\right\rfloor.
\]
This supply constraint may delay \texttt{T}/\texttt{T}$^\dagger$ gates until sufficient states have been produced.
Our aggregate model assumes no initial stock, omits factory startup time, and does not credit production during inserted walks.

To deliver these states, we allocate concentric routing rings around the $H\times W$ computation grid, expanding its dimensions to $(H+2F)\times(W+2F)$.
Each factory connects to magic-state tiles at row ends through a dedicated ring, with ports arranged to avoid crossing delivery paths (\cref{fig:magic-supply}a).
The delivery pass detects route conflicts but does not resolve them or account for transport delays.
Including both the delivery rings and factory footprints, the total space--time volume is
\[
  V_{\mathrm{chip+factory}}
  =\big[(H+2F)(W+2F)+FH_fW_f\big]T_{\mathrm{exec}},
\]
where $T_{\mathrm{exec}}$ counts execution steps, including walks.

\begingroup\setlength{\emergencystretch}{1em}
\smallskip\noindent\textbf{Magic-State Cultivation.}
Unlike distillation, cultivation~\cite{gidney2024magic} prepares magic states directly in the designated magic-state tiles, eliminating the need for external factories and delivery routes (\cref{fig:magic-supply}b).
However, preparation is stochastic because unsuccessful postselection requires a restart.
We therefore adapt routing to sampled state arrival times: a \texttt{T}/\texttt{T}$^\dagger$ gate waits until a state becomes available, while ready \CX{} gates may proceed independently.
Once prepared, a state remains at its tile until consumed, after which a new cultivation attempt begins.
\par\endgroup

\section{Mapping Initialization and Refinement}
\label{sec:appendix-mapping}

The placer in \cref{sec:mapping} initializes qubit positions using spectral placement and refines them through simulated annealing.
Since placement affects both routing contention and the space retained after compaction, we investigate how sensitive the final volume is to the initial assignment and the amount of refinement.
Extending the placement-effort study in \cref{sec:eval-sensitivity-main}, we compare seven initialization methods across different annealing budgets.
As required by \cref{sec:form-lanes}, the resulting virtual placement remains fixed throughout execution.

\begin{figure}[t]
  \centering
  \includegraphics[width=\columnwidth]{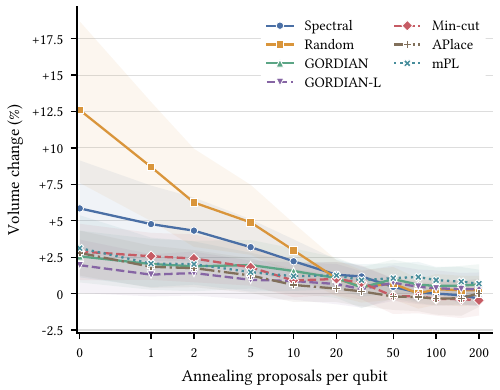}
  \caption{Mapping refinement on 56 circuits, showing geometric-mean volume changes relative to spectral at $100n$ proposals. Shading shows pointwise 95\% bootstrap intervals over circuits, and the horizontal axis is logarithmic above one and linear from zero to one.}
  \label{fig:mapping-budget}
  \Description{One plot compares spectral, random, GORDIAN, GORDIAN-L, Min-cut, APlace, and mPL initialization at zero, 1, 2, 5, 10, 20, 30, 50, 75, 100, 150, and 200 proposals per qubit. The horizontal axis is logarithmic above one and linear between zero and one. Markers show geometric-mean volume changes from spectral at 100 proposals per qubit, and shaded bands show pointwise 95 percent confidence intervals. Colors, markers, and line styles distinguish the seven methods.}
\end{figure}

\begingroup\setlength{\emergencystretch}{1em}
\smallskip\noindent\textbf{Initial Placements.}
We compare spectral and random initialization with five classical VLSI placement methods adapted to our virtual site array.
Spectral initialization uses the qubit-interaction graph, while random initialization assigns qubits uniformly to distinct sites.
Among the VLSI methods, GORDIAN minimizes quadratic wirelength with region-centroid constraints and recursive bisection~\cite{kleinhans1991gordian}.
GORDIAN-L modifies its objective to approximate linear wirelength~\cite{sigl1991analytical}.
Min-cut recursively partitions the array using balanced graph cuts~\cite{breuer1977mincut}.
APlace optimizes smooth wirelength and density penalties~\cite{kahng2005analytic}, while mPL uses multilevel coarsening and refinement of the interaction graph~\cite{chan2000multilevel}.
We implement the core formulations of these methods for comparison.
\par\endgroup

To isolate the effects of placement, all methods use the same demand-predicted array shape, fixed qubit orientations, and identical routing, compaction, walk, and on-demand magic-state supply models.
We independently verify each compiled program and average results over three random seeds per circuit, except for deterministic initializations without annealing, which require only one run.
Each annealing budget uses a complete cooling schedule.

\begin{table}[t]
  \caption{Mapping initialization on 59 circuits, where SA denotes $100n$ annealing proposals for $n$ qubits. Values are geometric-mean volume changes in percent relative to spectral with SA, with $\pm$ giving the bootstrap standard error in percentage points.}
  \label{tab:mapping-init}
  \centering\small
  \setlength{\tabcolsep}{4pt}
  \begin{tabularx}{\columnwidth}{@{}lRR@{}}
    \toprule
    \rowcolor{tlplum!15}
    Initial placement & No SA & With SA \\
    \midrule
    Spectral & $+5.61$ & $0.00 \pm 0.00$ \\
    Random & $+12.08$ & $+0.28 \pm 0.51$ \\
    GORDIAN & $+1.96$ & $+0.08 \pm 0.59$ \\
    GORDIAN-L & $+1.87$ & $-0.27 \pm 0.55$ \\
    Min-cut & $+3.18$ & $-0.17 \pm 0.54$ \\
    APlace & $+2.87$ & $-0.44 \pm 0.52$ \\
    mPL & $+3.16$ & $+0.92 \pm 0.52$ \\
    \bottomrule
  \end{tabularx}
\end{table}

As shown in \cref{tab:mapping-init}, refinement with $100n$ proposals brings the geometric-mean volumes of all seven initializers within 1\% of the annealed spectral baseline.
Most improvements occur early: just $30n$ proposals reduce the initial volume penalties of 2.0\%--12.6\% to 0.2\%--1.2\% (\cref{fig:mapping-budget}).
Increasing the budget from $100n$ to $200n$ further reduces spectral volume by only 0.34\%.
These results suggest that moderate annealing largely eliminates differences between initializers, with diminishing returns from additional refinement.

\section{Pauli-Based Computation}
\label{sec:appendix-pbc}

The execution model in \cref{sec:form-execution} routes each \CX{} and \texttt{T} gate individually.
Pauli-based computation (PBC) instead absorbs Clifford operations into Pauli rotations associated with \texttt{T} gates~\cite{bravyi2016pbc,litinski2019game}.
This can eliminate routed Clifford operations, but the resulting rotations may involve multiple data qubits, increasing routing demand.
To explore this tradeoff, we extend our compiler to support multi-qubit Pauli rotations through merges connecting multiple patch boundaries.
We then study how limiting the number of data qubits per merge affects space--time volume.

\begin{figure}[!t]
  \centering
  \includegraphics{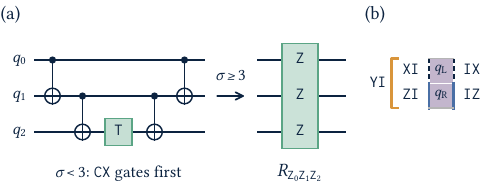}
  \caption{(a) A rotation about $\mathtt{Z}_0\mathtt{Z}_1\mathtt{Z}_2$ replaces four \CX{} gates and one \texttt{T} when the net capacity $\sigma$ is at least three. (b) Two-qubit patch boundaries, with $\mathtt{Y}$ using both $\mathtt{X}$ and $\mathtt{Z}$ contacts.}
  \label{fig:pbc-overview}
  \Description{Panel a shows a three-qubit circuit with CNOT zero to one, CNOT one to two, T on qubit two, and the two CNOTs in reverse order. Beside it, one rotation acts on the axis Z zero Z one Z two. Net capacities below three force the deferred CNOTs to execute before the T gate, while capacities of at least three allow the single rotation on three qubits. Panel b shows one two-tile patch encoding a left and a right qubit, labeled q L and q R vertically inside the purple patch. Left contacts are labeled XI and ZI, and right contacts IX and IZ, with factors ordered by the left and right qubits. An amber bracket joins the left contacts to indicate YI.}
\end{figure}

\begin{figure}[!t]
  \centering
  \includegraphics{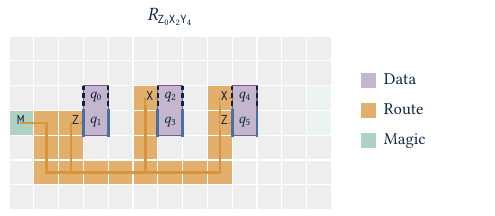}
  \caption{Routing a rotation about $\mathtt{Z}_0\mathtt{X}_2\mathtt{Y}_4$. A tree with a horizontal trunk joins the contacts of qubits 0, 2, and 4 and a magic state. Qubit 4 joins both contacts for its $\mathtt{Y}$ factor. }
  \label{fig:pbc-routing}
  \Description{A Pauli-rotation tree on a grid of three two-qubit patches. Data patches are purple, with labels q zero above q one, q two above q three, and q four above q five inside each patch. Routing tiles are amber and magic-state tiles green. Pauli labels sit outside the contacted patch boundaries. The tree touches Z on qubit zero, X on qubit two, and both X and Z on qubit four, and a horizontal trunk joins the vertical branches and a magic state on the left.}
\end{figure}

\smallskip\noindent\textbf{Capacity-Limited Transformation.}
PBC pushes Clifford gates to the end of the circuit, transforming each \texttt{T} gate into a Pauli rotation $R_P=e^{-i\pi P/8}$, followed by a final Clifford circuit~\cite{bravyi2016pbc,litinski2019game}.
Specifically, a \texttt{T} gate on qubit $q$ preceded by Clifford gates $C$ becomes a rotation about $P=C^\dagger\mathtt{Z}_qC$, where $P$ is a signed tensor product of $\mathtt{I}$, $\mathtt{X}$, $\mathtt{Y}$, and $\mathtt{Z}$.
For example, \cref{fig:pbc-overview}a shows how four \CX{} gates and one \texttt{T} gate can be replaced by a single rotation about $\mathtt{Z}_0\mathtt{Z}_1\mathtt{Z}_2$.
Although this transformation eliminates routed Clifford gates, the resulting rotations may involve more qubits and require more routing space.

To control this tradeoff, we represent each multi-qubit rotation as a routing net and introduce a net capacity $\sigma$, which limits the number of data qubits connected by a single net.
When absorbing additional Clifford gates would produce a rotation exceeding this capacity, we instead route the deferred \CX{} gates and restart the transformation with an empty Clifford part.
Thus, $\sigma=1$ restricts rotations to single qubits, while $\sigma=\infty$ allows unrestricted Clifford absorption.

For circuits with measured outputs, the final Clifford circuit can also be absorbed into the $\mathtt{Z}$-basis measurements~\cite{litinski2019game}, producing multi-qubit Pauli readouts that require routing but no magic states.
We apply the same capacity constraint to these readouts. If a readout exceeds $\sigma$, we execute the final Clifford circuit instead.

\begin{figure}[!t]
  \centering
  \includegraphics[trim=0 6bp 0 3bp,clip]{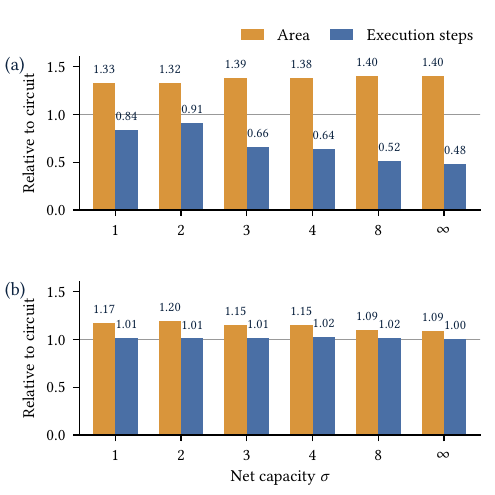}
  \caption{Area and execution steps relative to the circuit flow, in geometric mean over 29 circuits, with (a) on-demand magic states and (b) two distillation factories. Steps include walks, and area in (b) includes the 18 factory tiles. PBC uses more area but fewer steps on demand, while factories hold both flows to the same steps.}
  \label{fig:pbc-tradeoff}
  \Description{Paired bars show chip area in amber and execution steps in blue for net capacities sigma of one, two, three, four, eight, and infinity. Geometric-mean area ratios are 1.331, 1.324, 1.385, 1.379, 1.402, and 1.402. Execution-step ratios, including walks, are 0.838, 0.907, 0.663, 0.640, 0.517, and 0.482. A horizontal line at one marks the circuit reference. At every capacity, PBC increases chip area and reduces total execution steps in geometric mean.}
\end{figure}

\smallskip\noindent\textbf{Patches and Routing.}
To support PBC, we generalize the two-endpoint routes of \cref{sec:routing} to \emph{multi-pin nets}, where each pin represents a required patch-boundary contact.
An $\mathtt{X}$ or $\mathtt{Z}$ Pauli factor contributes the corresponding boundary contact, while a $\mathtt{Y}$ factor requires both but counts as only one data qubit toward the net capacity $\sigma$.
To expose these contacts, we represent each logical-qubit patch using two vertically adjacent tiles~\cite{litinski2019game}, with $\mathtt{X}$ and $\mathtt{Z}$ contacts on the upper and lower tiles, respectively (\cref{fig:pbc-overview}b).

We route each net as a connected tree spanning all required boundary contacts and, for Pauli rotations, a magic-state tile.
Candidate trees consist of a horizontal trunk along a routing lane with vertical branches to the contacts, as illustrated in \cref{fig:pbc-routing}.
If no candidate is available, we fall back to a greedy shortest-path search.
Route selection retains the criticality scores and lane costs of \cref{sec:routing}, together with the stage budgets of \cref{sec:layout}.
To prevent conflicting operations, each merge reserves both contacts of every participating qubit for its execution step.
We also preserve the order of rotations with anticommuting Pauli axes.

Finally, stage compaction extends the retained-coordinate rule of \cref{sec:routing} to include tree branch junctions, preserving connectivity between all pins.
Between stages, the walks of \cref{sec:form-lanes} move both tiles of each patch together.
An independent checker validates the resulting program.

\begin{figure*}[!t]
  \centering
  \includegraphics[width=\textwidth,trim=0 18bp 0 0,clip]{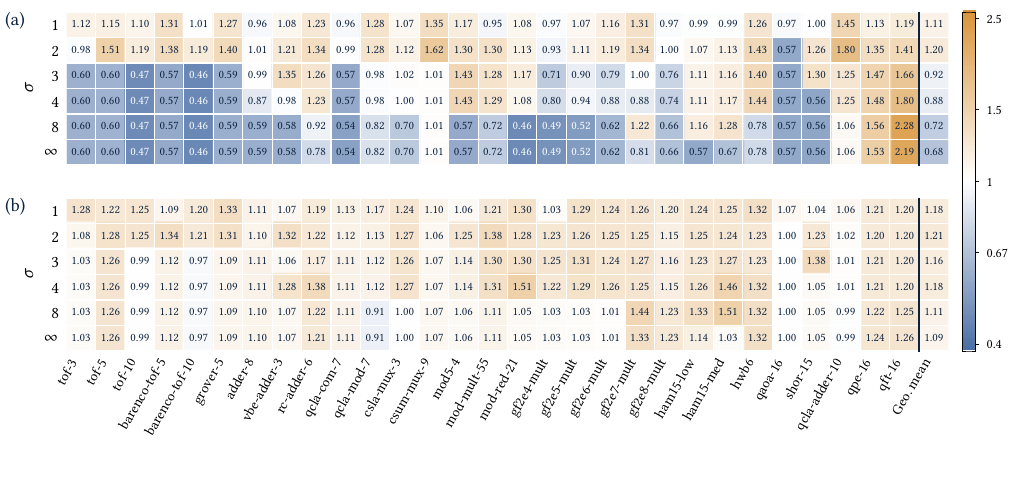}
  \caption{Volume relative to the circuit flow on 29 circuits across net capacities $\sigma$, with (a) on-demand magic states and (b) two distillation factories, and geometric means in the last column. Colors use a logarithmic scale centered at one. Capacities of three or more reduce geometric-mean volume on demand, but no capacity does with factories.}
  \label{fig:pbc-results}
  \Description{A six-row heatmap covering all 29 circuits and a final geometric-mean column, one row per net capacity sigma of one, two, three, four, eight, and infinity. The geometric means are 1.115, 1.201, 0.919, 0.883, 0.724, and 0.676. Blue indicates volume below the circuit baseline and amber indicates volume above it. QFT, QPE, qcla-adder-10, and csum-mux-9 are worse for every tested capacity, while several Toffoli circuits improve sharply at capacity three.}
\end{figure*}

We compare the circuit flow of \cref{sec:method} against PBC with net capacities $\sigma\in\{1,2,3,4,8,\infty\}$ on 29 circuits containing 5--36 qubits.
For each configuration, we use seed 0 and select the lowest-volume result among the three smallest-area shapes and the most square shape.
As in \cref{sec:eval}, we initially assume on-demand magic-state supply, with a fresh state available at every magic-state tile at the end of each patch row in every step.
We measure tile-step volume using \cref{sec:form-objective}, including walks and counting one step per routed \CX{}, Pauli rotation, or multi-qubit readout.
The model omits magic-state preparation, single-qubit Clifford costs, and classical-control delays.

Under on-demand magic-state supply, PBC reduces execution steps at the cost of larger chip area.
For $\sigma\ge3$, the step savings outweigh the area overhead in geometric mean (\cref{fig:pbc-tradeoff}a).
At $\sigma=\infty$, PBC reduces geometric-mean volume by 32.4\%, improving all tested Toffoli constructions and finite-field multipliers, as well as 9 of 11 adder and modular-arithmetic circuits (\cref{fig:pbc-results}a).
These circuits contain roughly one \texttt{T} gate per \CX{}, whereas QFT and QPE contain approximately four.
Clifford absorption therefore eliminates a larger fraction of routed operations in the former circuits without changing the \texttt{T} count.
In contrast, QFT and QPE incur 53\%--119\% higher volume despite shorter execution schedules, because their wider Pauli rotations require more routing space.
Even among arithmetic circuits, qcla-adder-10 and csum-mux-9 exhibit higher volume at every capacity.

The benefits of PBC diminish when magic-state production limits execution.
To evaluate this regime, we repeat the sweep with two $3\times3\times11$ distillation factories from \cref{sec:appendix-magic}.
Since PBC preserves the \texttt{T} count, both flows face the same total magic-state demand and production throughput.
Consequently, PBC's reduced routing depth provides little execution-time benefit: its step count remains between 1.00 and 1.02 times that of the circuit flow across all capacities (\cref{fig:pbc-tradeoff}b).
Meanwhile, its larger chip footprint increases geometric-mean volume by 9.2\%--20.7\% (\cref{fig:pbc-results}b).
These results indicate that PBC is most beneficial when routing, rather than magic-state production, limits execution.

\end{document}